\documentclass[reprint, superscriptaddress, amsmath,amssymb,prx]{revtex4-2}
\usepackage[utf8]{inputenc}
\usepackage[T1]{fontenc}
\usepackage{graphicx}
\usepackage{dcolumn, lipsum,graphicx,float}
\usepackage{diagbox}
\graphicspath{ {./Figures/} }
\usepackage{slashbox}
\usepackage{gensymb}
\usepackage{wrapfig}
\usepackage[dvipsnames]{xcolor}
\usepackage{svg}

\usepackage{hyperref}
\usepackage[modulo,pagewise]{lineno}
\usepackage{braket}
\usepackage{blkarray}
\usepackage{dirtytalk}
\usepackage[normalem]{ulem}
\newcommand{\figpanel}[2]{Fig.~\hyperref[#1]{\ref*{#1}(#2)}}
\newcommand{\figpanels}[3]{Fig.~\hyperref[#1]{\ref*{#1}(#2)-(#3)}}
\newcommand{\figspanels}[3]{Figs.~\hyperref[#1]{\ref*{#1}(#2)-(#3)}}

\newcommand{\figpanelNoPrefix}[2]{\hyperref[#1]{\ref*{#1}(#2)}}

\newcommand{\figurepanel}[2]{Figure~\hyperref[#1]{\ref*{#1}(#2)}}
\newcommand{\figurepanels}[3]{Figure~\hyperref[#1]{\ref*{#1}(#2)-(#3)}}

\newcommand{\figurepanelNoPrefix}[2]{\hyperref[#1]{\ref*{#1}(#2)}}
\newcommand{\figurespanels}[3]{Figures~\hyperref[#1]{\ref*{#1}(#2)-(#3)}}

\hypersetup{
    colorlinks=true,
    linkcolor=blue,
    citecolor = blue,
    urlcolor=blue
}

\usepackage{soul}
\usepackage{xcolor}

\begin{document}

\title{Mitigation of frequency collisions in superconducting quantum processors}
\email{amr.osman@chalmers.se}

\author{Amr Osman}

\affiliation{Department of Microtechnology and Nanoscience, Chalmers University of Technology, 412 96 Gothenburg, Sweden}
\author{Jorge Fernández-Pendás}
\affiliation{Department of Microtechnology and Nanoscience, Chalmers University of Technology, 412 96 Gothenburg, Sweden}

\author{Christopher Warren}
\affiliation{Department of Microtechnology and Nanoscience, Chalmers University of Technology, 412 96 Gothenburg, Sweden}

\author {Sandoko Kosen}
\affiliation{Department of Microtechnology and Nanoscience, Chalmers University of Technology, 412 96 Gothenburg, Sweden}
\author{Marco Scigliuzzo}
\affiliation{Institute of Physics, Swiss Federal Institute of Technology, Lausanne (EPFL), Lausanne, Switzerland}
\affiliation{Center for Quantum Science and Engineering, EPFL, Lausanne, Switzerland}

\author{Anton Frisk Kockum}
\author {Giovanna Tancredi}
\affiliation{Department of Microtechnology and Nanoscience, Chalmers University of Technology, 412 96 Gothenburg, Sweden}

\author {Anita Fadavi Roudsari}
\affiliation{Department of Microtechnology and Nanoscience, Chalmers University of Technology, 412 96 Gothenburg, Sweden}

\author{Jonas Bylander}
\affiliation{Department of Microtechnology and Nanoscience, Chalmers University of Technology, 412 96 Gothenburg, Sweden}

\begin{abstract}
The reproducibility of qubit parameters is a challenge for scaling up superconducting quantum processors. Signal crosstalk imposes constraints on the frequency separation between neighboring qubits. 
The frequency uncertainty of transmon qubits arising from the fabrication process is attributed to deviations in the Josephson junction area, tunnel barrier thickness, and the qubit capacitor.
We decrease the sensitivity 
to these variations by fabricating larger Josephson junctions and reduce the wafer-level standard deviation in resistance down to 2\%. 
We characterize 32 identical transmon qubits and demonstrate the reproducibility of the qubit frequencies with a 40~MHz standard deviation (i.e.~1\%) with qubit quality factors exceeding 2 million. 
We perform two-level-system (TLS) spectroscopy and observe no significant increase in the number of TLSs causing qubit relaxation. 
We further show by simulation that for our parametric-gate architecture, and accounting only for errors caused by the uncertainty of the qubit frequency, we can scale up to 100 qubits with an average of only 3 collisions between quantum-gate transition frequencies, assuming 2$\%$ crosstalk and 99.9$\%$ target gate fidelity.
\end{abstract}

\maketitle

\section{introduction}

Several physics and engineering challenges appear when one attempts to scale up quantum processors to a large number of qubits. 
In superconducting quantum technology, these challenges include qubit decoherence~\cite{richardson_2020, Muller2019, burnett_decoherence_2019}, circuit integration~\cite{rosenberg2017, Kosen2022}, and signal wiring~\cite{krinner2019}, to name a few examples.
%
%
Additionally, the choice of quantum processor architecture imposes requirements on the device fabrication tolerances, so that parameters such as the qubit frequencies match the designed target values within defined bounds. 
A careful design enables selective control and readout of each qubit with minimum signal crosstalk.
However, the current qubit-frequency reproducibility is hardly sufficient for scaling up beyond tens of qubits~\cite{MIT_reproducibility_2015, Osman2019, osman2021, kreikebaum2019improving,hertzberg2021, Verjauw2022}. This is particularly important in architectures with fixed-frequency qubits; for example, the cross-resonance gate architecture requires less than 6~MHz standard deviation of qubit frequencies in order to scale beyond 1000 qubits~\cite{hertzberg2021}. Simulation results~\cite{Morvan2022} show that ac-Stark-shift-based qubit architectures can be scaled up to 1000 qubits by deploying post-processing techniques such as laser annealing of Josephson junctions (JJs)~\cite{hertzberg2021}. 

Improved reproducibility of qubit frequencies was recently achieved through fabrication processes optimization~\cite{kreikebaum2019improving, osman2021, Verjauw2022}.
The uncertainty in the transition frequency of the commonly used transmon qubit~\cite{koch_charge-insensitive_2007} is caused by deviations in the normal-state resistance, $R_N$, of its Josephson junction and in the charging energy of the qubit capacitor, $E_C$. 
In a previous work, we showed that the $R_N$ variation has a strong junction-size dependence~\cite{osman2021}: smaller junctions are more sensitive to lithographic error, which results in JJ area deviation. 

In this work, we report on a reduction in the transmon qubit frequency variation down to 40~MHz (i.e.~1\%). 
We achieved this by increasing the JJ area to $\sim0.33\times0.33~\mu$m$^2$, three times larger, compared to our previous reference~\cite{osman2021}, while we compensated for the resistance change by growing a thicker tunnel-barrier oxide. 
We measured the resistances of thousands of JJs at room temperature and found consistency with low-temperature characterization results of 32 identical, fixed-frequency transmon qubits. 
Moreover, our analysis shows that, at the achieved variation in the junction resistance, variations of the qubit capacitance contribute approximately equally to the qubit-frequency variation.


We studied the implication of increasing the junction area on the qubit relaxation time by means of spectroscopy of two-level-system (TLS) defects. In four tunable qubits with a total junction area of 0.352~$\mu$m$^2$, we found only one TLS residing in a JJ over a 1-GHz frequency span. 
%
By simulation, we then estimated the number of junction-TLSs in a quantum processing unit (QPU) comprised of 100 qubits with JJs of size $0.33\times0.33~\mu$m$^2$, and found that approximately one TLS would collide in frequency with a qubit.

\newdimen\origiwspc%
\origiwspc=\fontdimen2\font

\fontdimen2\font=0.307em%
Finally, we simulated the qubit frequency distribution on a QPU architecture based on parametric gates (enabling iSWAP and CZ native gates) under the assumption of the frequency variation that we had achieved. 
Assuming 99.9\% target single and two-qubit gate fidelity, and 2\% signal crosstalk, we found an average of three frequency collisions on a 100-qubit QPU. While such collisions degrade the gate fidelity, circuit compilation might mitigate the effect.~Further reduction of the qubit-frequency variation to below 0.5\% would be required to achieve a high fabrication yield of collision-free QPUs.

\fontdimen2\font=\origiwspc

\section{JJ fabrication and resistance characterization at room temperature} \label{RT}

Our goal was to improve the reproducibility of the JJ fabrication process for transmon qubits by making larger junctions, while compensating for the increased area with a thicker tunnel barrier to target the same range of normal-state resistance $R_N$. 

\begin{figure*}
\centering
    \includegraphics[width=0.96\linewidth]{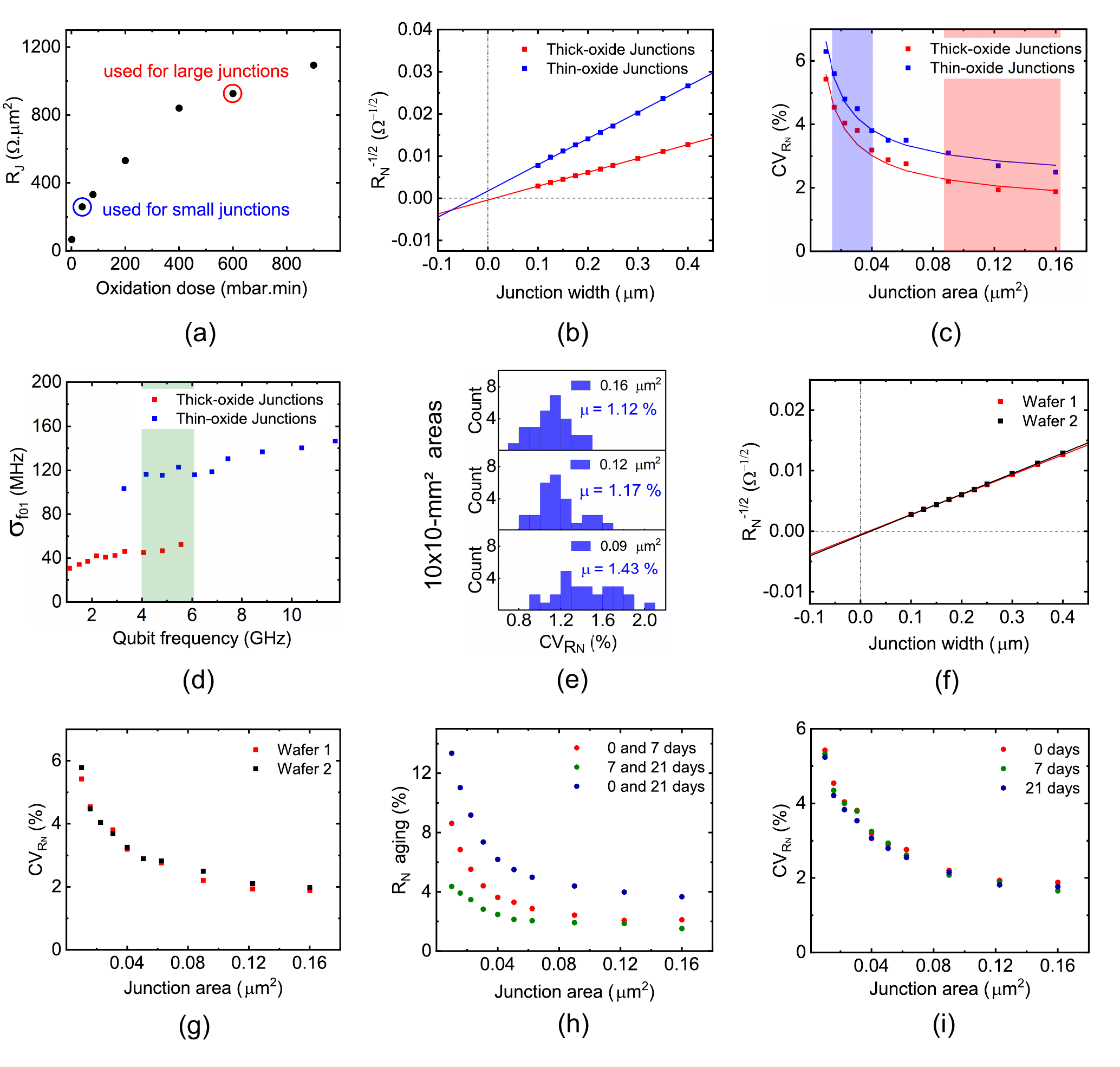}
    \caption{\textbf{Josephson junction resistance reproducibility.}\small{ (a) Inferred unit-area resistance vs.\@ oxidation dose. 
    (b) $R_N^{-1/2}$ vs.\@ nominal junction width $d$ for the two different oxidation conditions, with fits to Eq.~(\ref{eq:RN_vs_d}). 
    (c) Coefficient of variation of resistance vs.\@ junction area, with fits to Eq.~(\ref{cov}). The blue and red regions represent the junction areas for low and high oxidation dose, respectively, that result in qubit frequencies between 4 and 6~GHz. (d) Inferred standard deviation of qubit frequency [Eqs.~(\ref{eq:transmon})--(\ref{eq:transmon_stdev})] for the two different oxidation conditions showing an almost three-fold improvement for thicker-oxide junctions. 
    (e) Histograms of $CV_{R_N}$ for three junction sizes of interest over all possible combinations of $0.5 \times 0.5~\mathrm{mm}^2$ chips that result in $10\times10~\mathrm{mm}^2$ areas (27 possible combinations).
    (f) $R_N^{-1/2}$ vs.\@ $d$ for two different wafers with thick-oxide junctions. 
    (g) $CV_{R_N}$ for the wafers in (f). 
    (h) Aging of $R_N$ vs.\@ junction area.  
    (i) $CV_{R_N}$ vs.\@ junction area, showing no significant change over time. Note: the reference data points in blue in (b--d) are taken from Ref.~\cite{osman2021}. }}  
    \label{fig:fig1}
\end{figure*}

%
All JJs were fabricated using the patch-integrated Manhattan or cross-type (PICT) process~\cite{CMOS, osman2021} with areas between $0.1\times0.1$ and $0.4\times0.4~\mu$m$^2$. We first investigated how the oxidation conditions of the tunnel barrier---the oxidation time $t$ and pressure $p$---affect the unit-area resistance $R_J$. 
We tried seven different oxidation doses, defined as $p\times t$ (mbar$\,\cdot\,$min), and determined the corresponding $R_J$ values. The results are shown in \figpanel{fig:fig1}{a}.

The reference dose used for smaller-size junctions~\cite{osman2021}, 40 mbar$\,\cdot\,$min ($p = 2$~mbar and $t = 20$~min), yielded $R_J=240~\Omega~\mu$m$^2$. 
For the larger-area junctions, we selected an oxidation dose of 600~mbar$\,\cdot\,$min ($p =10$~mbar and $t =60$~min) to obtain a thicker tunnel barrier. 
An even higher oxidation dose would require larger junction sizes, beyond what the current resist-stack thickness can accommodate.
%

To characterize the variation in $R_N$ over a 50-mm wafer, we patterned 40 dies of size $5\times5\,$mm$^2$, each containing 100 test devices representing 10 junction sizes. 
We then measured $R_N$ of all junctions -- at room temperature and soon after fabrication -- using an automated, four-point probe station (see Appendix \ref{rn_meas}).

\figurepanel{fig:fig1}{b} shows the mean measured ${R_N}^{-1/2}$ as a function of the designed junction width $d$ for the new thick-oxide junctions (red, 600~mbar$\cdot$min) and a comparison with the reference junctions (blue, 40~mbar$\cdot$min~\cite{osman2021}). 
We fit to the model 
\begin{equation}
\label{eq:RN_vs_d}
    R_N = R_J/(d - \Delta d)^2, 
\end{equation}
where $\Delta d$ is the deviation (bias) from $d$ resulting from the lithography process. A positive (negative) $\Delta d$ indicates a smaller (larger) junction compared with the design. For the thick-oxide junctions we find $R_J=920~\Omega \cdot \mu$m$^2$ and $\Delta d= 16$~nm. (For the reference, $R_J=240~ \Omega \cdot \mu$m$^2$ and $\Delta d=-28$ ~nm.) We note that the difference in offset is only due to a change in the lithography process and is unrelated to the use of the thicker-oxide junctions.

\figurepanel{fig:fig1}{c} shows the coefficient of variation (CV) in $R_N$, $CV_{R_N}=\sigma_{R_N}/\overline{R_N}$, also referred to as the relative standard deviation, as a function of the junction area 
fitted to the equation~\cite{osman2021}
 \begin{equation}\label{cov}
\centering
 (CV_{R_N})^2 = (CV_{R_J})^2 (CV_{A})^2 + (CV_{R_J})^2 + (CV_{A})^2.
\end{equation}
$CV_{R_J}$ is due to the variation of the tunnel barrier thickness, while $CV_A = 2 \sigma_d / \sqrt{A}$ is the variation of the junction area due to the lithography process. For thick-oxide junctions, we extracted $CV_{R_J}=1.4\%$ and $\sigma_d=2.7$~nm, compared to $CV_{R_J}=1.8\%$ and $\sigma_d=4$~nm previously reported for the reference. 
We find that the uniformity of the tunnel barrier improved for the thicker-oxide junctions. However, we attribute the smaller $\sigma_d$ for the thicker-oxide junctions compared to that of the reference to a more careful descumming inspired by Ref.~\cite{kreikebaum2019improving}.
While this represents an improvement, the main benefit comes from fabricating larger-area junctions, enabled by the thicker tunnel-barrier oxide. 
The shaded regions in \figpanel{fig:fig1}{c} mark the junction sizes---and the corresponding $CV_{R_N}$ obtained---that are typically used for transmon qubits with frequencies within the 4--6-GHz range: blue for the thin-oxide, small junctions (reference) and red for the thicker-oxide, larger junctions. 

The transmon qubit frequency $f_{01}$ can be expressed in terms of $R_N$ and the charging energy $E_C$~\cite{koch_charge-insensitive_2007},
 \begin{equation}
\centering
    hf_{01} \approx \sqrt{\frac{2 \Delta \Phi_0}{e R_N} E_C} - E_C,
\label{eq:transmon}
\end{equation}
where $\Delta =$ 176 $\mu$eV is the superconducting gap of aluminum, $\Phi_0=h/2e$ is the magnetic flux quantum, $e$ is the electron charge, and $h$ is Planck's constant. Now assuming $E_C/h=200$\,MHz, we infer $f_{01}$ for every junction size from its measured $R_N$. The standard deviation is 
 \begin{equation}
    \sigma_{f_{01}} = 0.5 \, CV_{R_N} \,  f_{01},
\label{eq:transmon_stdev}
\end{equation}
since $CV_{f_{01}} = 0.5 \, CV_{R_N}$~\cite{Krantz2010InvestigationOT}, assuming no impact from the large qubit capacitor; an assumption we reconsider in the next section. 
We plot $\sigma_{f_{01}}$ as a function of $f_{01}$ in \figpanel{fig:fig1}{d}, where the green-shaded region indicates the qubit frequency range of interest. We find that $\sigma_{f_{01}} \approx 45$~MHz for qubits with thick-oxide JJs (red), compared to over 100~MHz for qubits with thin-oxide JJs (blue).

\begin{figure*}
\centering
    \includegraphics[width=0.95\linewidth]{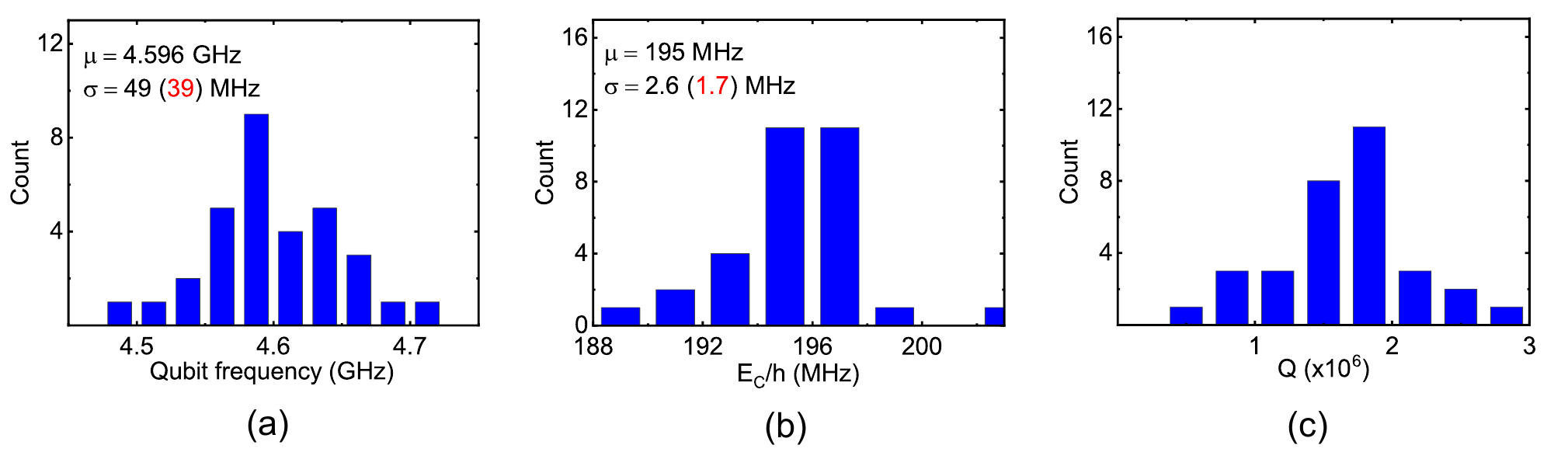}
    \caption{\textbf{Characterization of large-junction (thick-oxide) qubits.} \small{Histograms of (a) measured qubit frequencies, (b) charging energies, and (c) quality factors of 32 qubits. Values in red represent the standard deviations excluding three outliers.} }
    \label{fig:fig2}
\end{figure*}

This qubit-frequency variation is expected across the area of the 50-mm wafer. However, for QPUs with only dozens of qubits, the chip sizes may vary between 10$\times$10 and 15$\times$15~mm$^2$. Therefore, it is important to understand the qubit frequency variation over such areas. Here, we consider all the possible combinations of the neighboring 5$\times$5-mm$^2$ chips that result in chip areas of 10$\times$10~mm$^2$, which add up to 27 combinations. Then we extract $CV_{R_N}$ on each of these larger-area chips and plot them on a histogram [\figpanel{fig:fig1}{e}], where the average chip-level $CV_{R_N}$ for the smallest junction size of interest (0.09 $\mu m^2$), i.e., the highest $CV_{R_N}$, is 1.43$\%$.

To verify the results of our reproducibility study, we fabricated and characterized a replica of the thick-oxide wafer. \figurepanel{fig:fig1}{f} shows $R_N^{-1/2}$ as a function of $d$. The results are almost identical, with a shift of only $4\%$ in $R_N$ for junction sizes of interest. Most importantly, $CV_{R_N}$ is the same for both wafers [\figpanel{fig:fig1}{g}], indicating that the variation in $R_N$ is reproducible. We claim that a small shift in the mean resistance between two wafers, or on the same wafer due to aging, is tolerable, since that would shift all the qubit frequencies similarly. A $4\%$ shift in $R_N$, for example, would shift all the qubit frequencies by $2\%$. Assuming that the qubit frequencies on a QPU are distributed over 1 GHz (e.g.,~4--5~GHz), all qubits would shift by 80--100~MHz. The most crucial factor here is the reproducibility of $CV_{R_N}$ between wafers, since that determines how reliably we can fabricate QPUs without frequency collisions between qubits. 

Recent studies have shown that $R_N$ increases over time, in an ``aging'' process likely due to the redistribution of oxygen within the tunnel barrier, and that annealing of the junctions controllably accelerates it~\cite{Koppinen_2007, julin2016, Bilmes_2021}. We conducted a short-time study of the junction aging by measuring $R_N$ at two additional points in time: 7 and 21 days after fabrication. \figurepanel{fig:fig1}{h} shows the relative increase in $R_N$ after 7 days (green), from 7 to 21 days (red), and the total period from 0 to 21 days (blue). The resistance of the junction sizes of interest increased by 4$\%$ over the whole period, with a smaller increase from day 7 to day 21 compared to the first 7 days, indicating that the aging slows down over this timescale. Notably, $CV_{R_N}$ did not change, suggesting that aging may have little effect on the variation of $R_N$ across the wafer [\figpanel{fig:fig1}{i}].

\section{Characterization of the qubit frequency variation} \label{LT}
To compare the $f_{01}$ variation inferred from room-temperature $R_N$ measurements to the actual qubit frequency variation, we fabricated 32 fixed-frequency, uncoupled transmon qubits on a $14\times10~\mathrm{mm}^2$ area and measured their transition frequencies. All qubits were designed to have the same frequency, with the same $0.33\times0.33~\mu$m$^2$ junction area. \figurepanel{fig:fig2}{a} shows a histogram of the measured $f_{01}$ with a mean of 4.6~GHz and $\sigma_{f_{01}}=49$~MHz ($1.1\%$), which drops to 39~MHz ($0.8\%$) if we exclude three outliers. This is consistent with the 1.43\% average chip-level variation in resistance shown in \figpanel{fig:fig1}{e}. 

However, at such a low variation in $R_N$, we must consider charging-energy variations, presumably resulting from small deviations in the lithography and etching of the qubit capacitance. \figurepanel{fig:fig2}{b} shows the histogram of inferred $E_C$ values with $\sigma_{E_C}=1.7$~MHz (0.8$\%$), which contributes to the $f_{01}$ variation. Here, we extracted $E_C$ from the measured qubit spectrum, using the Hamiltonian of the Cooper-pair box. From these values of $\sigma_{f_{01}}$ and $\sigma_{E_C}$ we estimate $\sigma_{R_N}\approx 1\%$ in these qubits. This means that variations in $E_C$ and $R_N$ contribute approximately equally to the qubit frequency variation, and care should be taken to improve the wiring-layer lithography. 
%


A histogram of the quality factors $Q$ of the qubits is shown in \figpanel{fig:fig2}{c}. The mean $Q$ is 1.7$\times$10$^6$, which increases to 2.7$\times$10$^6$ if we account for the Purcell decay due to the readout resonator (see Appendix~\ref{ap_freq_meas}).

\section{Qubit coherence and TLS spectroscopy} \label{STLS}
JJs are known to host strongly coupled TLSs, with drastic negative effects on qubit coherence if their frequency is close to the qubit frequency. The density of TLS defects was reported to be $\rho_d\sim1.5~\mathrm{GHz}^{-1} \mu\mathrm{m}^{-2}$ for a 2-nm thick dielectric~\cite{TLS_KIT, Bilmes2022}. 
The thick-oxide junctions in this work are about three times larger than the reference, thin-oxide, smaller ones, and we therefore expect three times as many TLSs within these junctions. 
A statistical study of many qubits is out of the scope of this work; however, we performed a TLS-spectroscopy experiment on four tunable qubits made with asymmetric superconducting quantum interference devices (SQUIDs): two with large junctions and two with small.

\begin{figure*}
\centering
    \includegraphics[width=0.74\linewidth]{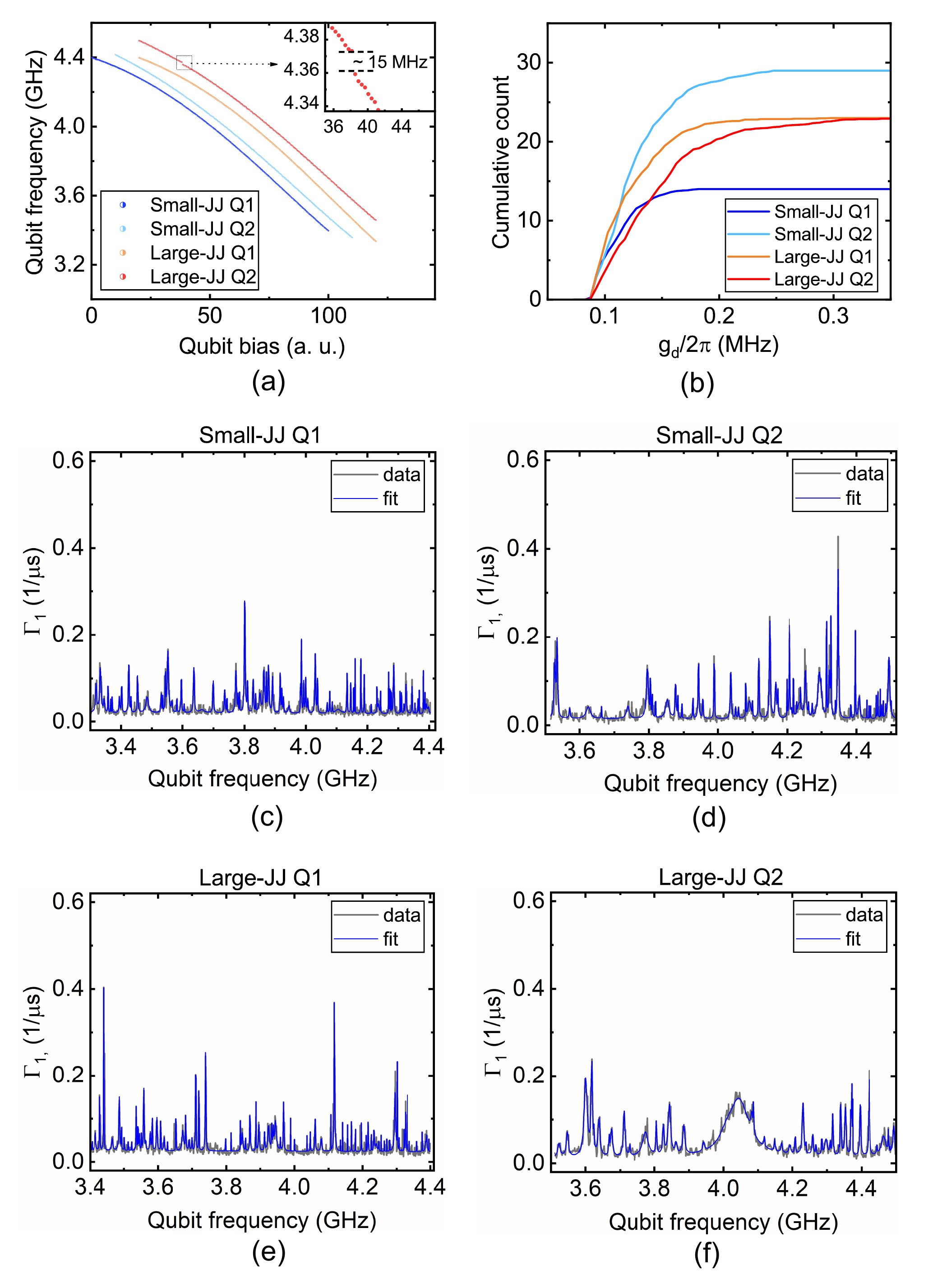}
    \caption{\small{\textbf{TLS spectroscopy}. (a) Measured frequencies vs.\@ qubit flux bias. Inset: a 15-MHz level-splitting due to a junction TLS interacting with one qubit. (b) Cumulative distribution of the detected TLSs in panels (c--f) with a coupling strength $g_d/2\pi > $ 90~kHz. (c--f) Relaxation rates of the four qubits over a 1-GHz span, where the peaks in the relaxation rate are due to the interaction between the qubit and different TLSs. Each peak is fitted to a Lorentzian (\ref{eq:lorentz}).
    (We suspect that the broad peak at 4.05~GHz in (f) originates from the same TLS shown in the inset of (a), but measured during a different cooldown.)}}
    \label{fig:fig3}
\end{figure*}

\begin{table}\label{squids}
\begin{ruledtabular}
\caption{\small{\label{tab:squids} TLS spectroscopy on tunable qubits with small (thin-oxide) and large (thick-oxide) JJs. Note that the total junction area is the design value, $d^2$, while the rest of the parameters are measured.}}
\centering
\footnotesize
\begin{tabular}{lccccc}

\shortstack{Qubit \\ ~} & \shortstack{Total \\ JJ area \\ ($\mu \mathrm{m}^2$)} & \shortstack{$f_{{01}_{max}}$ \\ (GHz) } & \shortstack{$T_1$ \\($\mu$s)}  & \shortstack{Number of \\ junction TLSs \\ (per GHz)} & \shortstack{Number of \\ non-junction TLSs \\ ($g_d > 90$~kHz)}\\\hline
Small-JJ Q1 & 0.044 & 4.467 & 45  &  0 & 14\\
Small-JJ Q2 & 0.044 & 4.617 & 79  & 0 & 29\\
Large-JJ Q1 & 0.132 & 4.495 & 50 & 0 & 23\\
Large-JJ Q2 & 0.132 & 4.666 & 65 & 1 & 23\\
\end{tabular}
\end{ruledtabular}

\end{table}

\begin{figure*}
\centering
    \includegraphics[width=1\linewidth]{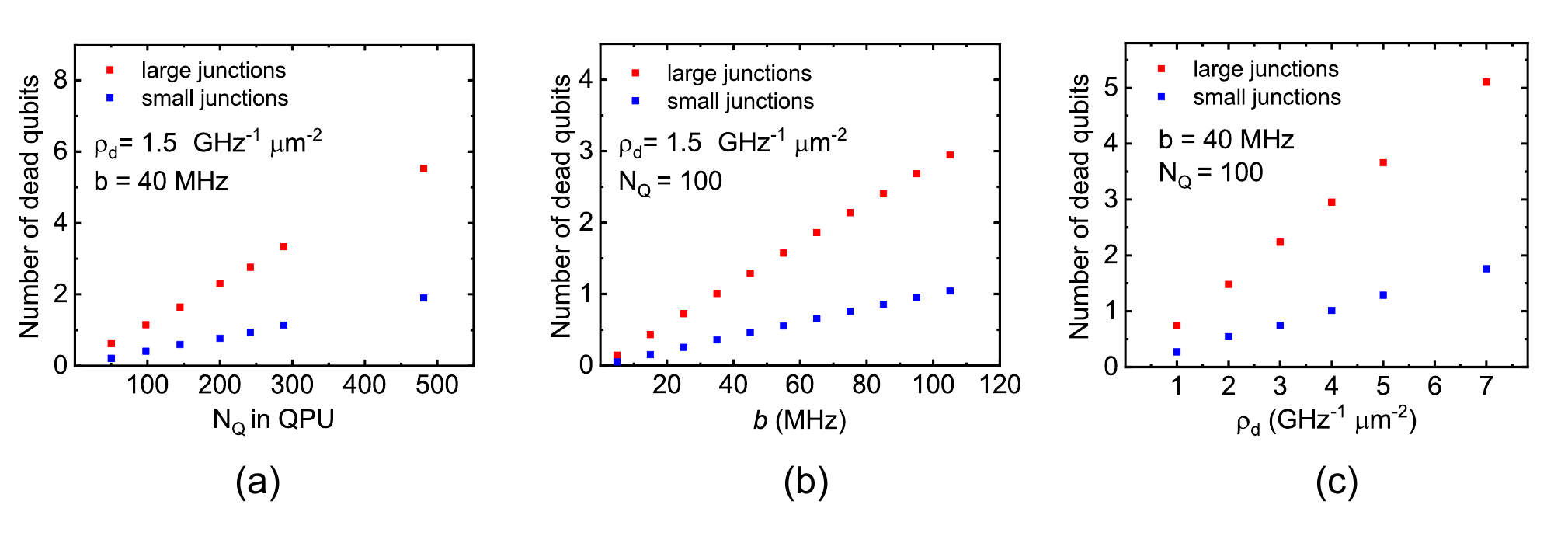}
    \caption{\small{\textbf{Scaling of TLS--qubit frequency collisions on a quantum processor}. Simulations showing the number of dead qubits due to junction TLS vs.\@ (a) number of qubits $N_Q$,  (b)  qubit--TLS collision bound $b$, and (c) TLS density $\rho_d.$ }}
    \label{fig:fig4}
\end{figure*}

Each small-junction qubit has a SQUID comprised of two JJs with areas 0.012 and 0.032~$\mu$m$^2$, and each large-junction qubit has JJ areas 0.035 and 0.097~$\mu$m$^2$. 
Table~\ref{tab:squids} shows the measured parameters of the four qubits. Assuming the literature value of $\rho_d$, 
we expect, on average, 0.13 and 0.40 junction TLSs in the small and large-junction qubits, respectively, and 0.53 junction TLSs in total. 
\figurepanel{fig:fig3}{a} shows the frequencies of the four qubits as a function of the qubit bias. Only one of the qubits (Large-JJ Q2) showed a signature of a strongly coupled TLS, with 15~MHz splitting (inset of \figpanel{fig:fig3}{a}), indicating that the TLS resides in a junction. 

We performed TLS-swap spectroscopy~\cite{barends_coherent_2013} on each of the four qubits to determine its relaxation rate $\Gamma_1$ as a function of the qubit frequency over a 1-GHz span. 
The cumulative result is shown in \figpanel{fig:fig3}{b} (with qubit-TLS coupling strength $g_d/2\pi > $ 90 kHz, limited by the resolution of the measurements), averaged over 29 measurements for each qubit. (See Appendix~\ref{ap_tls_spec} for more details.)
This was determined by analyzing the swap-spectroscopy data of each qubit [\figspanels{fig:fig3}{c}{f}]. Each peak is fitted to a Lorentzian,
\begin{equation}
    \Gamma_1 = \frac{2\; g_d^2 \; \Gamma }{\Gamma^2 + \delta^2} + \Gamma_{1,Q},
\label{eq:lorentz}
\end{equation}
where $g_d$ is the TLS--qubit coupling strength, $\Gamma$ is their total decoherence rate, and $\Gamma_{1,Q}$ is the qubit relaxation rate. $\delta = f_{01} - f_d$ is the detuning, where $f_d$ is the TLS (defect) frequency~\cite{barends_coherent_2013}. 

In \figpanel{fig:fig3}{b}, we observe little difference in the number of TLSs between the two types of qubits: the cumulative count ranges from 14 to 29 detectable TLSs for the four different qubits. \figurepanel{fig:fig3}{f} shows the signature of one strongly coupled TLS at 4.05~GHz (\emph{cf.}\@ the expectation value of 0.53 TLSs). 
This TLS most likely resides within the tunnel barrier, whereas the other, narrower Lorentzians are the signatures of TLSs that are located in the oxide surrounding the qubit electrodes~\cite{barends_coherent_2013}. 

\begin{figure*}
\centering
    \includegraphics[width=0.85\linewidth]{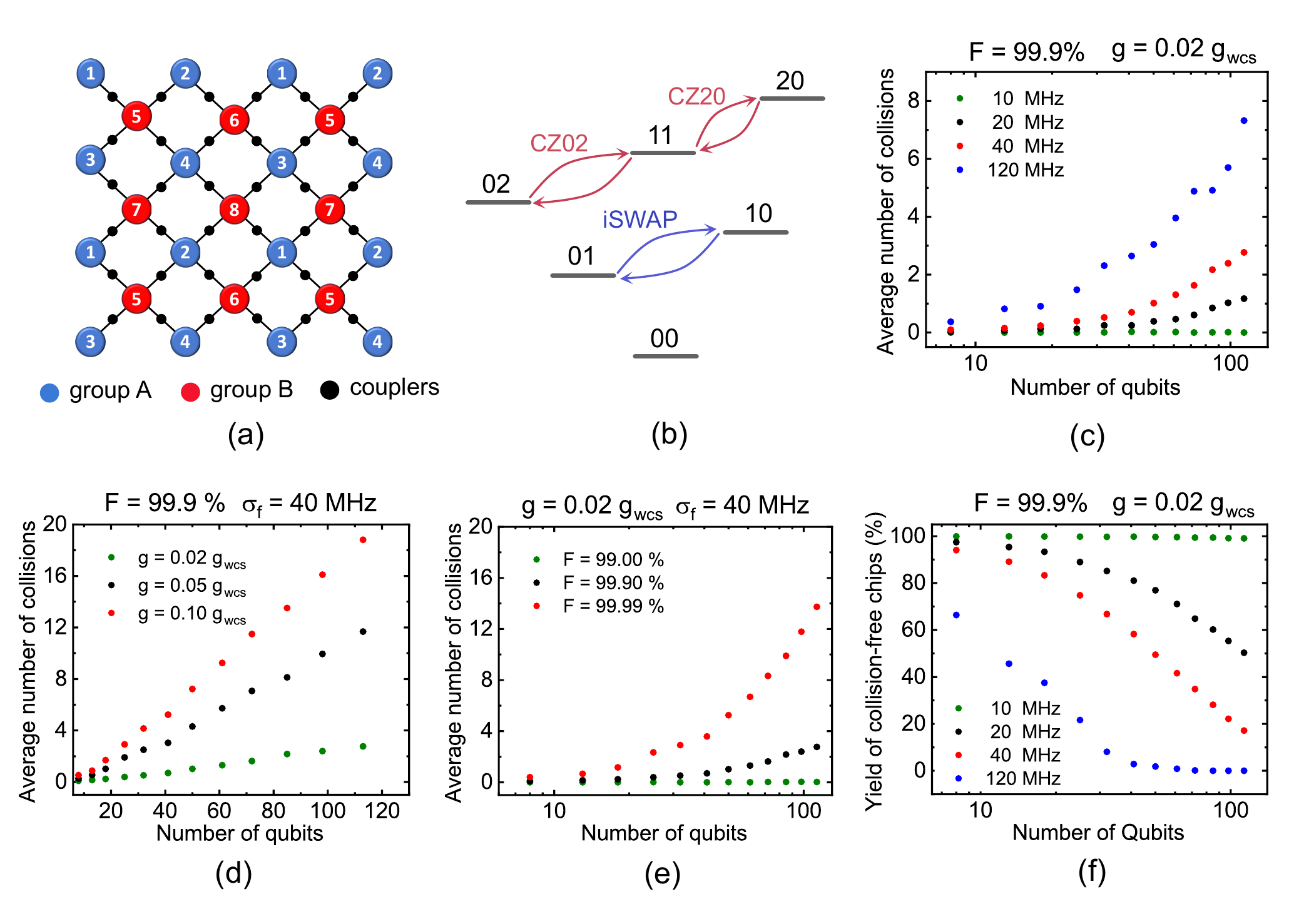}
    \caption{\small{ \textbf{Frequency collisions in a QPU architecture with fixed-frequency transmon qubits and two-qubit iSWAP and CZ gates driven by parametric couplers.} (a) A square lattice where qubits are divided into groups A (blue) and B (red), and tunable couplers in black. (b) Schematic of the two-qubit manifold, and the CZ02, CZ20 and iSWAP gates driven on their coupler. 
    (c--e) Simulation results showing the average number of collisions vs.\@ QPU size for (c) different qubit-frequency deviations with a fixed target gate fidelity and signal crosstalk level; 
    (d) various crosstalk levels with a fixed qubit-frequency deviation and target gate fidelity; 
    and (e) different target gate fidelities with a fixed signal crosstalk level and qubit-frequency deviation. 
    (f) Yield of collision-free QPUs for various qubit-frequency deviations with fixed target gate fidelity and signal crosstalk}.}
    \label{fig:fig5}
\end{figure*}

We proceeded to simulate the number of collisions between a qubit and a junction TLS in a QPU architecture comprised of fixed-frequency qubits connected on a square lattice [\figpanel{fig:fig5}{a}]. In this architecture, the qubits are assigned one of eight different frequencies over a span of 1~GHz (Section \ref{Qcollisions}). We ran a Monte Carlo simulation and randomly assigned a number $N_d^J$ of junction TLS defects over $N_Q$ qubits,
\begin{equation}\label{TLS_QPU}
        N_d^J = \rho_d \;  N_Q \; A_J,
\end{equation}
where $A_J$ is the JJ area (0.036 and 0.109 $\mu$m$^2$ on average for small and  large-junction qubits, respectively). We distributed these $N_d^J$ TLS frequencies uniformly over the 1~GHz span. A qubit is considered dead if its frequency is closer than $b$ to that of a junction TLS. 
\figurepanel{fig:fig4}{a} shows the number of dead qubits as a function of $N_Q$ for the literature value of $\rho_d$ and for $b=40$~MHz, which is similar to the high coupling strength $g_d$ between the qubit and a junction TLS~\cite{TLS_KIT}. As expected, the number of dead qubits on a large-junction QPU is about three times higher than on a small-junction QPU. 
However, only one qubit is expected to collide with a junction TLS on 100-qubit QPU with large junctions. We ran the simulation again, for 100 qubits, but with different values of $b$ to account for different average coupling strengths. \figurepanel{fig:fig4}{b} shows that the number of dead qubits scales linearly with $b$ within 100 MHz. Finally, \figpanel{fig:fig4}{c} shows the scaling with $\rho_d$ for $b=40$~MHz. Process improvements to reduce $\rho_d$ would increase the yield on a multi-qubit QPU.

\section{Frequency collision in a QPU} \label{Qcollisions}

In the following, we simulate the probability of frequency collisions between gate transitions in a quantum processor architecture based on a square lattice of fixed-frequency qubits and two-qubit gates driven by  parametric modulation of tunable couplers, see \figpanel{fig:fig5}{a} and Appendix \ref{app_col}. This architecture supports a gate set including iSWAP and CZ gates shown in \figpanel{fig:fig5}{b} on the two-qubit manifold~\cite{Roth_2017, Mckay2016, DiCarlo2009, Reagor2018}. The qubits are divided into two frequency groups: group A (blue), and group B (red); two-qubit gates are implemented between pairs of qubits in different groups by using the coupler (black). 
We can ensure that nearby resonances are separated as far as possible by carefully allocating the qubit frequencies and anharmonicities, under the constraints set by the available microwave signal bandwidth and required anharmonicity. Table~\ref{tab:crowding} shows our frequency $f_{01}$ and anharmonicity $\alpha$ allocation for the 8 qubits in the two frequency groups in \figpanel{fig:fig5}{a}. In the ideal scenario, where the frequency of the implemented qubits is exactly the same as the design value, crosstalk has almost no effect on gate fidelities. However, in reality, qubits deviate from their design parameters, and signal crosstalk may inadvertently drive a non-target resonance, degrading those fidelities~\cite{Abrams2019}.\\

\begin{table}\label{crowding}
\begin{ruledtabular}
\caption{\small{\label{tab:crowding} Qubit frequency $f_{01}$ and anharmonicity $\alpha$ allocation in a square lattice [\figpanel{fig:fig5}{a}] that avoids collision between neighboring single and two-qubit gates in a quantum processing unit based on the parametric-gate architecture.}}
\centering
\small
\begin{tabular}{lcccccccc}

\shortstack{Qubit} & 1 & 2 & 3 & 4 & 5 & 6 & 7& 8 \\\hline \shortstack{\\$f_{01}$\\ (GHz)} & 4.3 & 4.404 & 4.508 & 4.612 & 4.988 & 5.092 & 5.196 & 5.3 \vspace{0.2cm}\\\hline  \shortstack{\\$\alpha$ \\ (MHz) } & -156 & -156 & -156 & -156 & -260 & -260 & -260 & -260 \\

\end{tabular}
\end{ruledtabular}
\vspace{-5mm}
\end{table}

We performed a Monte Carlo simulation, randomly assigning qubit frequencies with a mean equal to the target frequency (Table \ref{tab:crowding}) and different standard deviations $\sigma_f$. We then counted the number of collisions found in the lattice. A collision is triggered if two neighboring single or two-qubit gates are closer to each other than a certain bound. This bound depends on two parameters: AC crosstalk $X_{AC}$ and target gate fidelity $F$, where the higher they are, the more stringent (larger) those bounds become. However, collisions between gates on the same coupler have a different mechanism. Derivations of collision bounds are found in Appendix~\ref{app_col}. Here, we define $g = X_{AC} \, g_{wcs}$, where $g$ is the gate strength on one qubit if a neighboring qubit is driven, and $g_{wcs}$ is the worst-case scenario of $g$, when $X_{AC}$ is 100$\%$. \figurepanel{fig:fig5}{c} shows the average number of collisions on an $N$-qubit QPU for $\sigma_f = $ 10, 20, 40, and 120~MHz, where the last two cases of $\sigma_f$ represent the large and small-junction qubits, respectively [\figpanel{fig:fig1}{d}].
Here we assume $X_{AC} = 2\%$ and $F = 99.9\%$. The average number of collisions on a 100-qubit QPU is 2.7 times lower for the large-junction qubits introduced in this work, compared to the reference small-junction qubits. We then consider the case $\sigma_f = 40$~MHz and simulate the average number of collisions in the presence of greater crosstalk (2, 5, and 10\%), showing a 7-times increase in the number of collisions if the crosstalk increases to 10$\%$ [\figpanel{fig:fig5}{d}]. \figurepanel{fig:fig5}{e} shows that changing the threshold of the required fidelity has a significant effect, where the number of collisions increases five times for $F= 99.99\%$. 
Finally, we find that in order to obtain a high yield of collision-free QPUs, the standard deviation in qubit frequency must drop to 10--15~MHz [\figpanel{fig:fig5}{f}].

We note that, in this simulation, we neglected the effect of qubit decoherence~\cite{Abad_2022, Abad_2023}, imperfection of the control pulse, driving further than nearest-neighboring gate, and simultaneous driving of more than one neighboring qubit gate. We do consider that only one of the two available CZ gates needs to be implemented on the QPU, in which case we neglect collisions between the iSWAP gate and the other CZ gate.

\section{conclusion}

We demonstrated frequency reproducibility within about 40~MHz (standard deviation) for fixed-frequency transmon qubits on a wafer---without any post-fabrication adjustment---a more than two-fold improvement compared to our previous reference qubits. This was achieved through fabricating larger-area Josephson junctions, which are less sensitive to lithographic variation, with thicker oxide to obtain the targeted junction resistance. 
At this sub-percent level, two terms contribute approximately equally: variation of the tunnel-junction resistance, due to the amorphous oxide barrier and the tunnel-junction area; and
variation of the qubit capacitance, due to lithography and etching inhomogeneity.

These larger-junction qubits showed good quality factors exceeding $2\times10^6$ and no noticeable increase in the number of strongly coupled two-level system defects adversely affecting quantum coherence. We predict through simulation that, at the achieved level of 40 MHz standard deviation in qubit frequency, the number of collisions on a 100-qubit system drops by threefold. However, a reduction of the frequency variation to about 15~MHz is needed for a high yield of collision-free chips and high-fidelity gate operations in our chosen quantum-processor architecture. 
This improvement can be achieved by further optimization of the lithography and oxide growth processes. 
Alternative architectures, e.g.,\@ combining fixed-frequency and tuneable qubits, may relax the constraints at the expense of added complexity. Signal crosstalk, on the other hand, can be minimized within this architecture by improved wire routing, grounding, and shielding.


\section{Acknowledgement}

This research was funded by the Knut and Alice Wallenberg Foundation through the Wallenberg Center for Quantum Technology (WACQT) and by the EU Flagship on Quantum Technology H2020-FETFLAG-2018-03 project 820363 OpenSuperQ. The authors acknowledge the use of Nanofabrication Laboratory (NFL) at Myfab Chalmers. \\

\appendix


\section{Junction resistance characterization} \label{rn_meas}
The room-temperature resistances, $R_N$, of the Josephson junctions were characterized using an automated probe station MPI-TS2000 with a four-point probe card. Resistance was measured in a current-biased mode by passing $-1\,\mathrm{to}\,+1\mu$A with 20 steps. Only measurements with coefficient of determination higher than 99$\%$ were chosen. Based on that, the measurement yield for all junction sizes on the large-junctions wafer (Fig.~\ref{fig:fig1} of the main text) is $> 99.4 \%$, except for junction sizes of 0.1225 $\mu$m$^2$ and 0.0625 $\mu$m$^2$. For these two sizes, two known faulty sets of probes reduced the measurement yield; however, this has virtually no effect on the results. On the same wafer, only 7 out of 3729 successful measurements were excluded for being outliers, giving a fabrication yield of 99.8 $\%$. All of those 7 junctions were considered open circuits.

\section{Qubit frequency characterization} \label{ap_freq_meas}

\begin{figure}

\centering
    \includegraphics[width=1\linewidth]{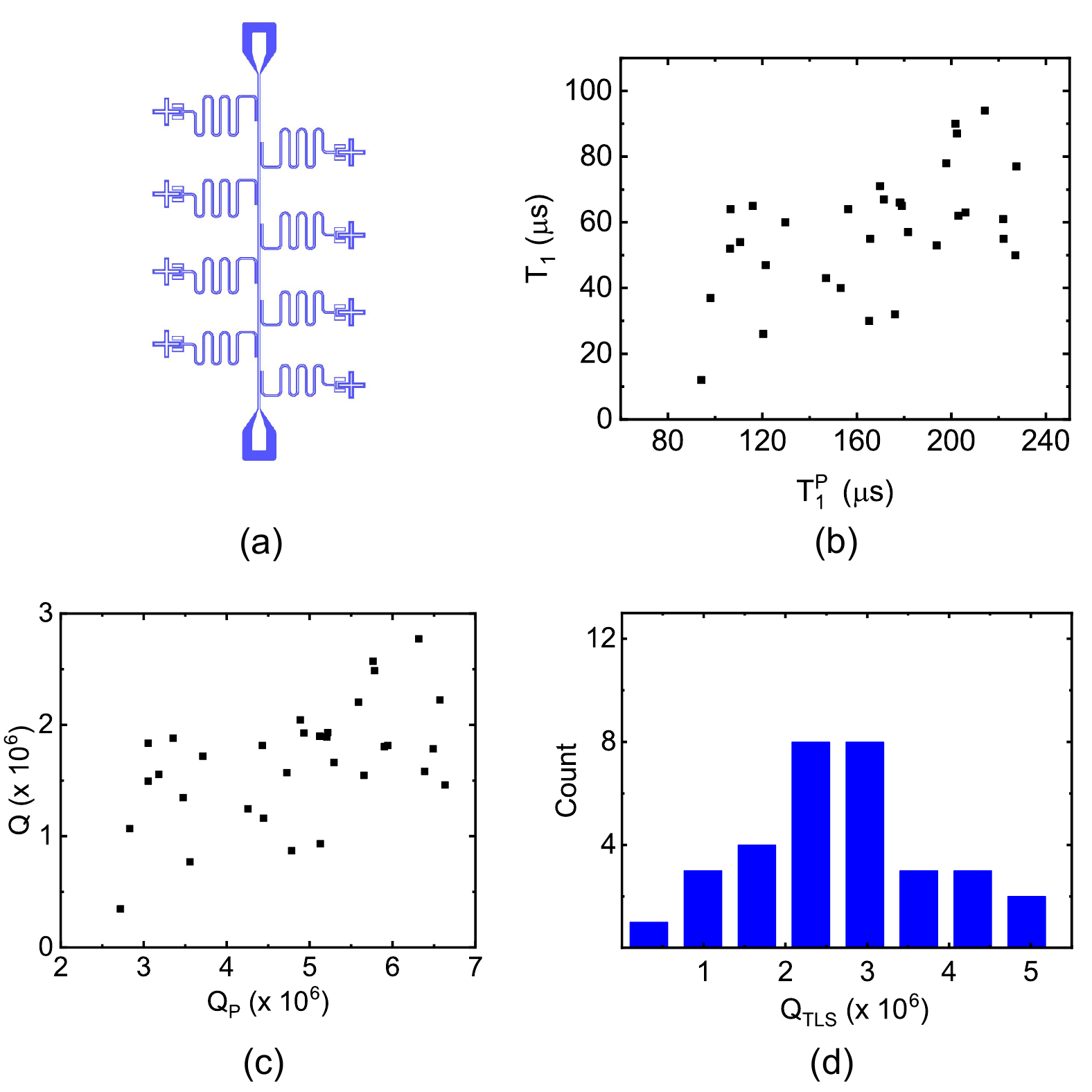}
    \caption{\textbf{Lifetime vs. Purcell limit of the characterized large-junction qubits with thick oxide.}\small{ (a) Layout of the 4 characterized chips. (b) Measured $T_1$ vs. the Purcell limit. (c) Measured quality factor vs. the quality factor limited by Purcell decay. (d) Histogram of the quality factors after accounting for the Purcell decay. }}
    \label{fig:fig6}

\end{figure}

We fabricated 32 qubits using the PICT process~\cite{osman2021}. Each set of 8 qubits was placed on a 5$\times$7~mm$^2$ die [\figpanel{fig:fig6}{a}], with four dies forming an equivalent rectangular die with 10$\times$14 mm$^2$ area. The average physical spacing between two vertically or horizontally neighboring qubits is 2.5 mm. We packaged each die in a copper box, and characterized the qubits at 10~mK during the same cooldown of a dilution refrigerator.

In the main text, we claim that some of the qubits with lower quality factor $Q$ [\figpanel{fig:fig2}{c}] are limited by Purcell decay. The Purcell limit on qubit lifetime, $T_1^P$, is given by $\Delta^2/(\kappa \, g_{01}^2)$, where $\kappa$ is the readout resonator linewidth, $\Delta$ is the detuning between the readout resonator and the qubit frequencies, and $g_{01}$ is their coupling strength~\cite{koch_charge-insensitive_2007}. We also consider a correction factor $\omega_r / \omega_q$, where $\omega_r$ and $\omega_q$ are the resonator and qubit frequencies, respectively~\cite{Jeffrey2014}. \figurespanels{fig:fig6}{b}{c} show the mean $T_1$ and $Q$ of every qubit, as a function of $T_1^P$ and $Q_P = \omega_q T_1^P$, respectively. The scaling of $T_1$ with $T_1^P$ shows that the relaxation rate had a contribution from Purcell decay. We then separate out the effect of Purcell decay from the the otherwise dominant decay due to TLS, using $1/Q = 1/Q_{TLS} + 1/Q_P$, and find the mean quality factor due to TLSs, $Q_{TLS}=2.7\times 10^6$.\\

\begin{figure*}

\centering
    \includegraphics[width=0.9\linewidth]{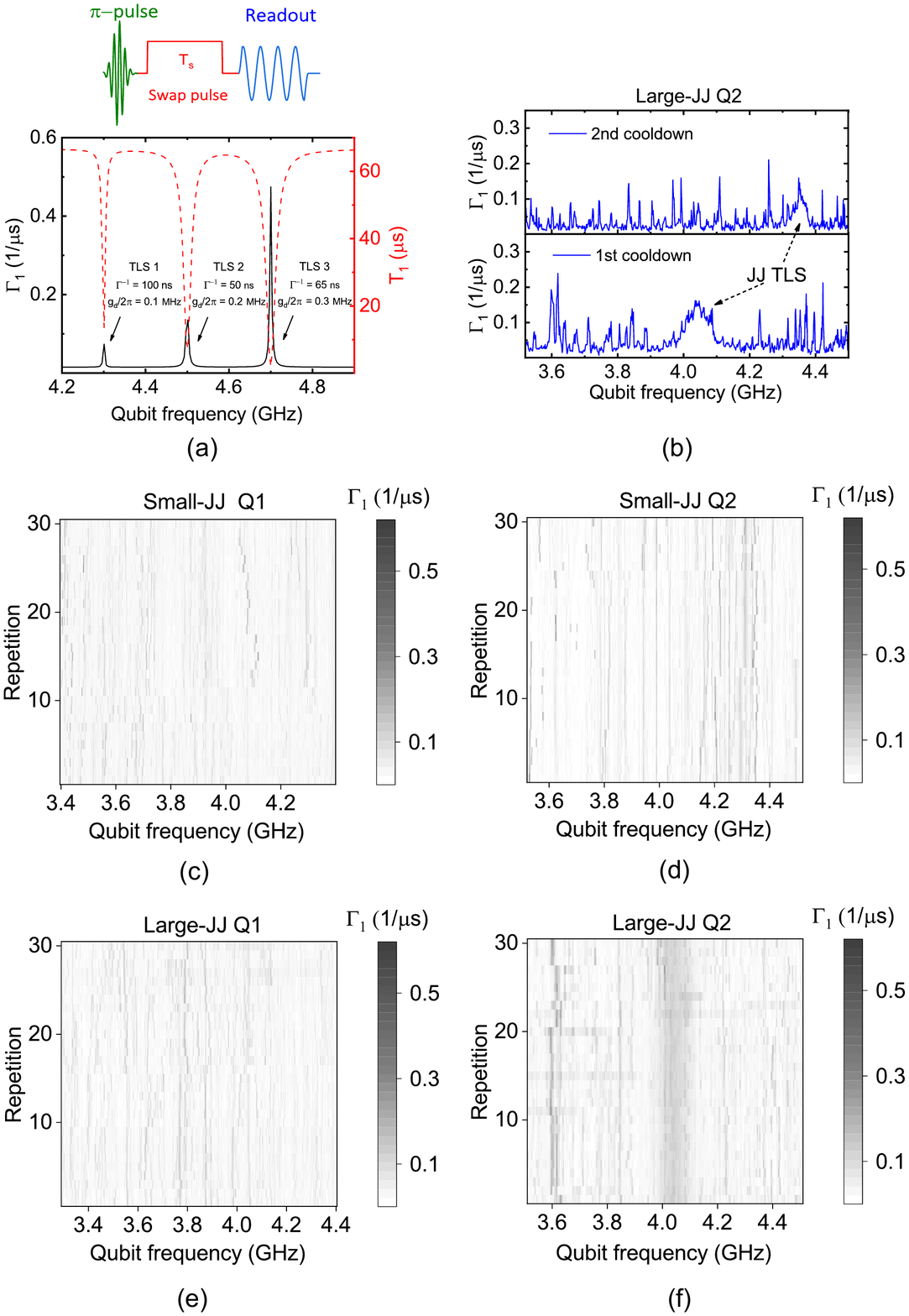}
    \caption{\textbf{TLS spectroscopy.} \small{(a) Pulse sequence for TLS swap spectroscopy. (b) Simulation of the effect of three different TLSs with different coupling strengths on the relaxation rate of the qubit. (c--f) Repeated TLS-spectroscopy measurements of the four qubits over an average time span of $\sim$15 hours for each qubit.}}
    \label{fig:fig7}

\end{figure*}

\section{TLS spectroscopy} \label{ap_tls_spec}
The four tunable qubits for which we characterized the TLS density were designed with asymmetric SQUIDs to reduce the sensitivity to flux noise, where the target asymmetry was 0.5 for each SQUID. Qubit Small-JJ Q1 is identical to qubit Large-JJ Q1 except for the junction sizes, and similarly for Small-JJ Q2 and Large-JJ Q2 (Table~\ref{tab:squids} of the main text).

To extract the qubit decay rate at different frequencies, we applied the sequence in \figpanel{fig:fig7}{a}-top, where a $\pi$-pulse excites the qubit to the $\ket{1}$ state, then a DC-flux pulse of fixed amplitude tunes the qubit frequency for some duration $T_s$, and then back, whereupon the qubit excited-state population $P$ is read out. By repeating the experiment while increasing the flux pulse duration in steps, at the same flux amplitude, we extract an exponentially decaying curve for $P$, with a time constant equal to the qubit lifetime at that frequency. Here, instead of sampling the population at different $T_s$ values, we use a 3-point method~\cite{TLS_KIT}. The three points are the $\ket{1}$-state population (i) $P_1$ at $T_s = 0$ , (ii) $P_s$ at $0 < T_s < T_1$ and (iii) $P_0$ when no $\pi$-pulse is applied. $T_1$ at the new frequency is then given by 

\begin{equation}
T_1 = T_s / \log{ \bigg( \frac{P_1 - P_0}{P_s - P_0}}\bigg).
\label{eq:3point}
\end{equation}

The advantage of the 3-point method is the shorter measurement time, which reduces the effect of temporal fluctuations in TLS frequencies, giving a more accurate estimation of TLS parameters: frequency $f_d$, coupling $g_d$, and decoherence rate $\Gamma_d$. Details on the $T_1$ extraction error of the 3-point method can be found in the supplementary material of Ref.~\cite{TLS_KIT}.\\ 

The effect of a collision between the qubit and a TLS on the qubit lifetime depends on their coupling strength and the qubit and TLS lifetimes. To illustrate how $T_1$ or $\Gamma_1$ behave as a function of qubit frequency, \figpanel{fig:fig7}{a}-bottom shows simulated $T_1$ data of a tunable qubit in the presence of 3 TLSs at 3 different frequencies, using Eq.~(\ref{eq:lorentz}) in the main text. \figurespanels{fig:fig7}{c}{f} show the relaxation rate $\Gamma_1$ as a function of frequency for the four tunable qubits mentioned in the manuscript, calculated directly from Eq.~(\ref{eq:3point}) [no fitting to Eq.~(\ref{eq:lorentz})]. For every qubit, the TLS-spectroscopy experiment was repeated $\sim$30 times over a time span of $\sim$15 hours. The resolution of these measurements varies from 0.7 to 1.5 MHz. Exemplary traces taken out of these figures, along with their fit, are shown in \figspanels{fig:fig3}{c}{f} of the main text. These heatmaps show that the detected TLSs for each qubit are persistent over all measurements, with some TLSs stable in their frequency and others showing telegraphic switching, which may be attributed to TLS-TLS interaction~\cite{Lisenfeld2015, Muller2019}.\\

\begin{figure}

\centering
    \includegraphics[width=1\linewidth]{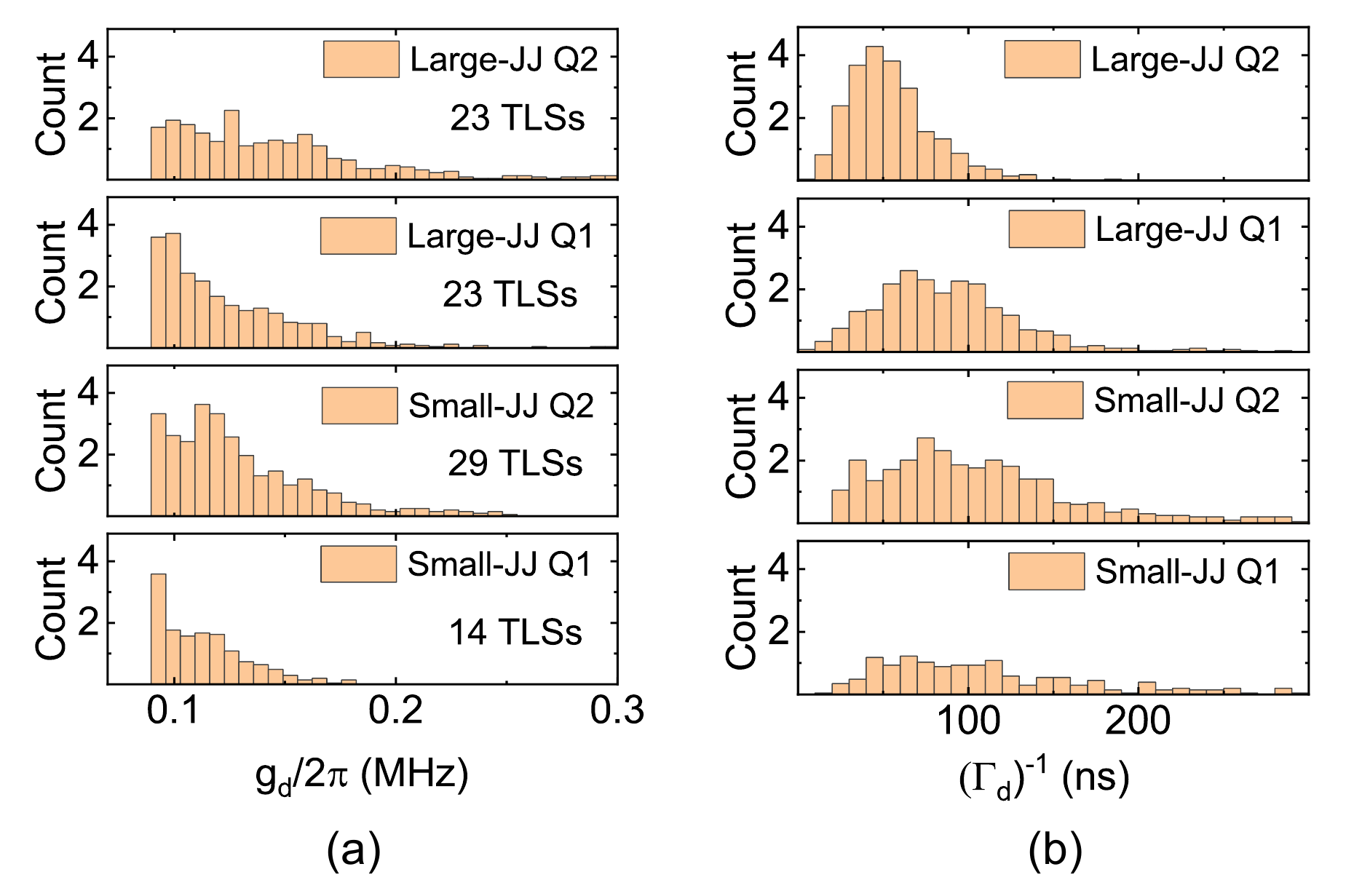}
    \caption{\textbf{Coupling strengths and decoherence time of TLSs.} \small{(a) Histograms of the qubit-TLS coupling strength $g_d$ for the four qubits. (b) Histograms of the extracted decoherence time $\Gamma_d^{-1}$ of TLSs in the 4 qubits. }}
    \label{fig:fig8}

\end{figure}

\figurepanel{fig:fig7}{b} shows TLS spectroscopy traces on qubit Large-JJ Q2 during two different cooldowns. We claim that the detected junction TLS in this qubit (indicated by the arrows) shifted in frequency between the two cooldowns, from 4.045 to 4.365 GHz. This claim is based on the signature of the TLS and the proximity in frequency. The level splitting of \figpanel{fig:fig3}{a} of the main text corresponds to this TLS in the second cooldown.\\

For a fair comparison between the four qubits when estimating the total number of detected non-junction TLSs [\figpanel{fig:fig3}{b} of the main text], only TLSs with coupling strength $g_d/2\pi >$ 90 kHz were considered. This is because the measurements of the four qubits were not equally sensitive to weakly coupled TLSs due to their different resolution. We find empirically that at $g_d/2\pi \gtrsim $ 90 kHz, qubits are equally sensitive to TLSs, where the distribution of $g_d$ behaves as expected (Supplementary material of Ref.~\cite{barends_coherent_2013}). Histograms of the distribution of $g_d/2\pi$ and $\Gamma_d^{-1}$ are shown in \figspanels{fig:fig8}{a}{b}. For each qubit, the values on these histograms are extracted from Lorentzian fits for all the TLSs detected in the repeated measurements [\figspanels{fig:fig7}{c}{f}], then normalized to the total number of TLSs in that qubit. This total number is taken from the trace with the highest number of fitted TLSs, which is 14, 29, 23 and 23 for qubits Small-JJ Q1, Small-JJ Q2, Large-JJ Q1 and Large-JJ Q2, respectively. TLS coherence is distributed between 50 and 200~ns, which matches previously reported values~\cite{barends_coherent_2013, TLS_KIT}. \\

\section{Parametric-gate-based QPU} \label{app_col}
\subsection{Frequency allocation}\label{freqalloc}

The two-qubit gates that we consider in this work are the iSWAP and CZ. The first one is implemented by driving the resonance between states $\lvert 10 \rangle$ and $\lvert 01 \rangle$ for a time such that the population of those two states is completely swapped, giving a $\pi / 2$-phase. The second is implemented by driving the resonance between $\lvert 11 \rangle$ and either $\lvert 20 \rangle$ or $\lvert 02 \rangle$ for a full Rabi cycle, in order to bring back all the population initially at $\lvert 11 \rangle$ onto itself with an acquired phase of $\pi$. The modulation frequency for a coupler between qubits $j$ and $k$ can be approximated as the difference of bare frequencies between their relevant states:
\begin{align}
  \omega_{\rm iSWAP}^{jk} =&\ \lvert \omega_{10} - \omega_{01} \rvert = \lvert f_{01}^{j} - f_{01}^{k} \rvert \, , \\
  \omega_{\rm CZ(20)}^{jk} =&\ \lvert \omega_{11} - \omega_{20} \rvert = \lvert f_{01}^{k} - f_{01}^{j} - \alpha^{j} \rvert \, , \\
  \omega_{\rm CZ(02)}^{jk} =&\ \lvert \omega_{11} - \omega_{02} \rvert = \lvert f_{01}^{j} - f_{01}^{k} - \alpha^{k} \rvert \, ,
\end{align}
where $f_{01}^{k}$ is the $k$th qubit's 01 transition frequency, and $\alpha^{k}$ is its anharmonicity,  $\alpha^{k} = f_{02}^{k} - 2 f_{01}^{k}$. These expressions indicate that detunings between coupled qubits, and their qubit anharmonicities, must be chosen with frequency crowding in mind. Moreover, for each pair of qubits, detuning alone sets the iSWAP frequencies and then each CZ frequency will be placed at a distance from the iSWAP frequency given by the anharmonicities of the qubit whose second excited state is temporarily populated. In particular, for transmons, which have negative-anharmonicity, the CZ will be placed above iSWAP if the qubit visiting the second excited state has a lower frequency than the other one, and below otherwise.

Under the assumption that all population before the application of a gate is within the computational states, we can state the problem of frequency crowding in the following way: no other resonance involving computational-space states should be in a similar range of frequencies as the modulation frequencies of the gates. This also includes resonances that involve states where the coupler is in its first excited state, and therefore it also constrains the choice of coupler DC bias at runtime, which is out of the scope of this study.

In this work, we consider a square lattice, like the one in \figpanel{fig:fig5}{a}. This means that each blue (red) qubit is coupled to four other red (blue) qubits, and each coupling is mediated by a coupler (black). Provided a fixed and relatively limited bandwidth, in order to have all important resonances accessible and equally spaced, we set each qubit in the square lattice to be surrounded by qubits equispaced in frequency. As a result, one ends up with two groups of qubits, i.e., qubits 1 to 4 (group A: blue), and 5 to 8 (group B: red). Each group has four similar but different frequencies and all four share the same anharmonicity. The relative value of the anharmonicities between the two groups is optimized to make the frequencies as equispaced as possible, resulting in a larger anharmonicity $\alpha_B$ for group B than $\alpha_A$ for group A. Additionally, this allocation scheme results in a high(low)-frequency-CZ group, where the qubits from group A (B) have their second excited state temporarily populated. Accordingly, we considered the following when allocating the frequencies and anharmonicities in Table~\ref{tab:crowding}.

\textbf{Bandwidth:} We allocate qubit frequencies within 1~GHz, limited by the available bandwidth of our control electronics for single and two-qubit gates. Here, we allow the high-frequency CZs to be outside the 1-GHz bandwidth, assuming we are interested only in the low-frequency CZs. Even in this case, only two of the high-frequency CZs are actually above 1 GHz.

\textbf{Qubit-qubit detuning and anharmonicity:} Here, we present the conditions that guarantee that physically neighboring iSWAP and CZ gates are as spectrally equispaced as possible:
\begin{itemize}
    \item Qubits within each group, A and B, are equispaced with $\Delta_Q$.
    \item $\alpha_A = $ 1.5 $\Delta_Q$ and $\alpha_B = $ 2.5 $\Delta_Q$.
\end{itemize}
Deciding on one of the values of $\Delta_Q$, $\alpha_A$, or $\alpha_B$, allows us to find the other two. Our allocation in Table~\ref{tab:crowding}, with $\Delta_Q = 104$~MHz, $\alpha_A = 156$~MHz, and $\alpha_B = 260$~MHz, guarantees that these conditions are satisfied while taking into account two additional constraints: the lowest low-frequency CZ gate is higher than 100~MHz, and the minimum $E_J/E_C$ is higher than 60.

The strategy explained above is not necessarily optimal, and it is possibly not the only logical procedure to arrive at a nearly equispaced frequency crowding. However, it is a simple and automatable procedure, and it allows us to scale up with the repetition of a simple unit cell consisting of qubits 1 to 8.

\subsection{Collision bounds}

In the following, we explain how far apart two resonances need to be in order to avoid collision in the frequency spectrum. While single-qubit gates are based on Rabi oscillations between the two computational states of a qubit, the parametric two-qubit gates typically work through the activation of Rabi oscillations between states in the two-qubit Hilbert space [\figpanel{fig:fig5}{b}], using the periodic oscillation of an additional coupling element. The two types of resonances are different in nature, resulting in two different crowding conditions: detunings between adjacent qubits are constrained by the single-qubit gates, while two-qubit parametric gates are constrained by the choice of parameters for each pair of coupled qubits, as well as for adjacent pairs of coupled qubits.
\begin{table*}
\centering\begin{tabular}{|c c c c c | l | l | l | l |}
  \hline
  Qb1 & C1 & Qb2 & C2 & Qb3 & Rabi states & Shared & Section \\\hline
  1qg & & 1qg & & & $\lvert 0xy \rangle$-$\lvert 1xy \rangle$, $\lvert x0y \rangle$-$\lvert x1y \rangle$ & None & \ref{sec:colliding_1qg} \\
  1qg & & & & 1qg & $\lvert 0xy \rangle$-$\lvert 1xy \rangle$, $\lvert xy0 \rangle$-$\lvert xy1 \rangle$ & None & \ref{sec:colliding_1qg} \\
  & CZ20 \ \ CZ02 & & & & $\lvert 11x \rangle$-$\lvert 20x \rangle$, $\lvert 11x \rangle$-$\lvert 02x \rangle$ & $\lvert 11x \rangle$ & \ref{sec:colliding_CZ_desired_CZ_same_coupler} \\
  & CZjk \ \ iS \ \ \ \ & & & & $\lvert 11x \rangle$-$\lvert jkx \rangle$, $\lvert 10x \rangle$-$\lvert 01x \rangle$ & None & \ref{sec:colliding_iSWAP_desired_CZ_same_coupler} \\
  & \ \ \ \ iS \ \ CZjk & & & & $\lvert 10x \rangle$-$\lvert 01x \rangle$, $\lvert 11x \rangle$-$\lvert jkx \rangle$ & None & \ref{sec:colliding_CZ_desired_iSWAP_same_coupler} \\
  & CZ20 & & CZ20 & & $\lvert 11x \rangle$-$\lvert 20x \rangle$, $\lvert x11 \rangle$-$\lvert x20 \rangle$ & $\lvert 111 \rangle$ & \ref{sec:colliding_CZ_desired_CZ_neighbor_coupler} \\
  & CZ20 & & CZ02 & & $\lvert 11x \rangle$-$\lvert 20x \rangle$, $\lvert x11 \rangle$-$\lvert x02 \rangle$ & $\lvert 111 \rangle$ & \ref{sec:colliding_CZ_desired_CZ_neighbor_coupler} \\
  & CZ02 & & CZ02 & & $\lvert 11x \rangle$-$\lvert 02x \rangle$, $\lvert x11 \rangle$-$\lvert x02 \rangle$ & $\lvert 111 \rangle$ & \ref{sec:colliding_CZ_desired_CZ_neighbor_coupler} \\
  & CZ02 & & CZ20 & & $\lvert 11x \rangle$-$\lvert 02x \rangle$, $\lvert x11 \rangle$-$\lvert x20 \rangle$ & $\lvert 111 \rangle$, $\lvert 020 \rangle$ & \ref{sec:colliding_CZ_desired_CZ_neighbor_coupler_more_states} \\
  & CZjk & & iS & & $\lvert 11x \rangle$-$\lvert jkx \rangle$, $\lvert x10 \rangle$-$\lvert x01 \rangle$ & $\lvert 110 \rangle$, $\lvert 201 \rangle$ & \ref{sec:colliding_iSWAP_desired_CZ_neighbor_coupler} \\
  & iS & & CZ20 & & $\lvert 10x \rangle$-$\lvert 01x \rangle$, $\lvert x11 \rangle$-$\lvert x20 \rangle$ & $\lvert 011 \rangle$ & \ref{sec:colliding_CZ_desired_iSWAP_neighbor_coupler} \\
  & iS & & CZ02 & & $\lvert 10x \rangle$-$\lvert 01x \rangle$, $\lvert x11 \rangle$-$\lvert x02 \rangle$ & $\lvert 011 \rangle$, $\lvert 102 \rangle$ & \ref{sec:colliding_CZ_desired_iSWAP_neighbor_coupler_more_states} \\
  & iS & & iS & & $\lvert 10x \rangle$-$\lvert 01x \rangle$, $\lvert x10 \rangle$, $\lvert x01 \rangle$ & $\lvert 010 \rangle$, $\lvert 101 \rangle$ & \ref{sec:colliding_iSWAP_desired_iSWAP_neighbor_coupler} \\\hline
  \end{tabular}

  \caption{Types of collisions in an iSWAP- and CZ-based parametric-gate architecture. The first column shows a list of the desired gates. The second column shows a corresponding list of states involved in the relevant Rabi oscillations. The third column shows a corresponding list of states that are simultaneously part of the desired and colliding gates. The last column includes references to subsections in Appendix~\ref{app_col}, explaining the collision process. In this table, $x,y=0,1,2...$, and $j,k=0,2$; 1qg means single-qubit gates; iS means iSWAP; CZ20 and CZ02 refer to CZ gates implemented using $\lvert 20 \rangle$ or $\lvert 02 \rangle$ as intermediate states, and Qbk and Ck stand for ``Qubit k'' and ``Coupler k'', respectively. Our convention for state identification in the table is $\lvert Qb1 \ Qb2 \ Qb3 \rangle$. }
  \label{tab:gates}

\end{table*}
\bigskip

The physics of Rabi oscillations are well-understood, and this can guide us in identifying potential crowding issues. There is partial population transfer even when there exists some detuning from the resonance, resulting in so-called detuned Rabi oscillations. They are typically faster than full Rabi oscillations and the amount of population that is transferred is smaller. However, it is precisely the fact that some population transfer happens when slightly detuned from the center of the resonance that makes the resonances have an effective width in frequency. In what follows, we calculate the appropriate frequency collision bounds from the width of the resonances.

Let us consider the different possible collisions schematically. We associate the different operations to the qubits and couplers in a linear chain, as it is the simplest model of a system containing qubits that are nearest and next-to-nearest neighbors, as well as couplers that are nearest neighbors. In Table~\ref{tab:gates}, we enumerate the types of collisions in an iSWAP- and CZ-based parametric-gate architecture. From the list of shared states between the Rabi oscillations associated to the desired and colliding processes, it becomes clear that there are two distinct situations, each one modeled by slightly different effective Hamiltonians: nearby resonances without shared states, and nearby resonances with shared states.

\textbf{Nearby resonances without shared states:} This happens when two relevant resonances of the system are near each other in frequency space but each one involves a different pair of states, without any state simultaneously in both pairs. In order to avoid effects from this kind of behavior, we need to ensure that, by driving a resonance at its center, i.e., without any detuning, the detuning to the nearby resonance is not low enough to effectively drive it. The colliding operation can be easily modeled by a two-level system, and the collision bound will be given by the value of detuning that produces an infidelity of the desired gate below some certain reasonable small threshold. The physics of a two-level system that is accidentally driven can be modeled with the following Hamiltonian:
\begin{equation}
  H_{2} = \left(
    \begin{array}{c c}
      \omega_{1} & g \cos \omega_{d} t \\
      g \cos \omega_{d} t & \omega_{2}
    \end{array}
  \right) \, ,
\end{equation}
where we have implicitly taken $\hbar = 1$, $g$ is the amplitude of the drive, $\omega_{d}$ is the driving frequency, and $\omega_{1}$ and $\omega_{2}$ are the energies of the two levels, respectively. In the rotating frame of the energy levels, the Hamiltonian reads
\begin{equation}
  H_{2}^{rf} = \left(
    \begin{array}{c c}
      0 & \frac{g}{2} e^{-i \Delta t}\\
      \frac{g}{2} e^{i \Delta t} & 0
    \end{array}
  \right) \, ,
\end{equation}
where $\Delta = \lvert \omega_{2} - \omega_{1} \rvert - \omega_{d}$ is the small detuning between the driving frequency and the resonance between the two levels, and we have implicitly applied the rotating-wave approximation by dropping the fast rotating terms of frequency equal to $\omega_{2} - \omega_{1} + \omega_{d}$. The time-evolution operator for this system, which allows to relate the initial state with the state at any time according to $\lvert \psi (t) \rangle = U (t) \lvert \psi (0) \rangle$,  can be analytically obtained:
\begin{widetext}
\begin{equation}
  U (t) = \left(
    \begin{array}{c c}
      e^{-i \frac{\Delta}{2} t} \left( \cos \frac{\Omega t}{2} + i \frac{\Delta}{\Omega} \sin \frac{\Omega t}{2} \right) & - i e^{-i \frac{\Delta}{2} t}\frac{g}{\Omega} \sin \frac{\Omega t}{2} \\
      -i e^{i \frac{\Delta}{2} t} \frac{g}{\Omega} \sin \frac{\Omega t}{2} & e^{i \frac{\Delta}{2} t} \left(\cos \frac{\Omega t}{2} - i \frac{\Delta}{\Omega} \sin \frac{\Omega t}{2} \right)
    \end{array}
  \right) \, ,
\end{equation}
\end{widetext}
where $\Omega = \sqrt{\Delta^{2} + \lvert g \rvert^{2}}$ is the usual Rabi frequency. Although we will be running our computations numerically, this analytical expression allows us to better understand the underlying physics of the colliding processes and, in the limit $\Delta \rightarrow 0$, the behavior of the desired gate operations.

\textbf{Nearby resonances with shared states:} This happens when two relevant resonances of the system are near each other in frequency and one of the states is part of both resonances. In order to avoid effects from this kind of behavior, we need to ensure that the detunings between the shared state and the other two states are sufficiently different to be able to turn on the resonances separately. This can be easily modeled by a three-level system, and the collision bound will be given by the values of detuning that give an infidelity of the desired gate below some certain reasonable small threshold. The physics of a periodically driven three-level system, such that the drive is aimed at coupling the first two levels but it also couples one of them to a third level, can be modeled with the following Hamiltonian:
\begin{equation}
H = \left(
  \begin{array}{c c c}
    \omega_{1} & g_{1} \cos \omega_{d} t & 0 \\
    g_{1} \cos \omega_{d} t & \omega_{2} & g_{2} \cos \omega_{d} t \\
    0 & g_{2} \cos \omega_{d} t & \omega_{3} \\
  \end{array}
\right) \, ,
\end{equation}
where $g_{j}$ ($j=1,2$) is the amplitude of the $j$th drive, $\omega_{d}$ is the driving frequency, and $\omega_{k}$ ($k=1,2,3$) are the energies of the three levels (where $\omega_{2} > \omega_{1}$). We can change to the rotating frame of the energy levels by a simple unitary transformation, resulting in the following Hamiltonian:
\begin{equation}
H_{rf} = \left(
  \begin{array}{c c c}
    0 & \frac{g_{1}}{2} & 0 \\
    \frac{g_{1}}{2} & 0 & \frac{g_{2}}{2} e^{-i \Delta t} \\
    0 & \frac{g_{2}}{2} e^{i \Delta t} & 0
  \end{array}
\right) \, ,
\end{equation}
where $\Delta = \lvert \omega_{3} - \omega_{2} \rvert - \omega_{d}$, the resonance between the first two states is hit perfectly ($\omega_{d} = \omega_{2} - \omega_{1}$), and we have implicitly applied the rotating-wave approximation by dropping the fast-rotating terms. The validity of this approximation might be under question if $g_{2}$ was comparable to $\lvert \omega_{3} - \omega_{2} \rvert$, but we typically stay out of this regime. The time-evolution operator of this system cannot be obtained analytically in simple terms.

In these models, one does not typically have control over the coupling strengths of the undesired processes. Thus, in our simulations we consider the possibility that they are equal to their worst-case-scenario value, which is in most cases simply equal to their coupling strength if they were activated on purpose, or to a fraction of it. Another underlying assumption that we need to make is that one activates at most one undesired resonance simultaneously with the desired one, or, equivalently, that one can analyze separately the different collisions for each particular process. This is a reasonable approximation, though, since one needs to avoid all collisions to make a certain gate have high fidelity.

The metric that we will use to measure the quality of the resulting gates is the average gate fidelity, which can be computed by the use of the following formula \cite{Pedersen2007}:
\begin{equation}
  F = \frac{\left\lvert \mathrm{Tr} \left( M U_{g}^\dagger \right) \right\rvert^2 + \mathrm{Tr} \left( M M^\dagger \right)}{d (d+1)} \, ,
\end{equation}
where $M$ is the propagator of the actual process, $U_{g}$ is the unitary of the ideal gate, and $d$ is the dimension of the computational space, i.e., $d = 2^n$ for $n$-qubit gates. Thus, an essential step in the following derivations is the description of the gates with errors due to undesired activation of other resonances using a propagator $M$. The two distinct types of collisions listed above are actually good descriptions for different possible collisions, both for single- and two-qubit gates. Therefore, we now move on to analyze on a case-by-case basis the impact of the collisions for the different gates using these simplified models.

\subsubsection{Colliding single-qubit gates: bounds for qubit detunings}
\label{sec:colliding_1qg}

\begin{itemize}
  \item Desired gate: $U_{g} = X \otimes \mathbb{I}$ in $t_{g} = 20 \, \mathrm{ns}$. We assume that other single-qubit gates will behave similarly.
  \item Gate's Rabi-oscillation coupling strength: $g_{g} = \pi / t_{g} = 25 \, \mathrm{MHz}$.
  \item Colliding process 

    \begin{align}
      H_{rf} =&\ \frac{g_{g}}{2} \left( \lvert 10 \rangle \langle 00 \rvert + \lvert 11 \rangle \langle 01 \rvert \right) \nonumber \\
      +& \frac{g}{2} e^{i \Delta t} \left( \lvert 01 \rangle \lvert 00 \rvert + \lvert 11 \rangle \langle 10 \rvert \right) + \mathrm{H.c.}
     \end{align}

  \item Worst-case coupling strength: as strong as the gate's coupling strength, ${g_{wcs} = 25 \, \mathrm{MHz}}$.
  \item Detuning (collision bound) as a function of coupling strength and fidelity threshold is shown in Table~\ref{table:single_qubit_gates}. These results give us, in principle, $\Delta$ for both nearest and next-nearest neighbors, although we expect the latter to have even less stringent conditions as the crosstalk should be weaker.
\end{itemize}

\begin{table}[ht]
  \centering
  \begin{tabular}{|l | l l l|}
    \hline
    & $F > 0.99$ & $F > 0.999$ & $F > 0.9999$ \\\hline
    $g = 0.1 g_{wcs}$
    & 0.022
    & 0.041
    & 0.047
    \\ 
    $g = 0.2 g_{wcs}$
    & 0.036
    & 0.046
    & 0.098
    \\ 
    $g = 0.3 g_{wcs}$
    & 0.041
    & 0.050
    & 0.198
    \\ 
    $g = 0.4 g_{wcs}$
    & 0.044
    & 0.098
    & 0.299
    \\ 
    $g = 0.5 g_{wcs}$
    & 0.047
    & 0.148
    & 0.449
    \\ 
    $g = 0.6 g_{wcs}$
    & 0.092
    & 0.200
    & 0.649
    \\ 
    $g = 0.7 g_{wcs}$
    & 0.095
    & 0.296
    & 0.898
    \\ 
    $g = 0.8 g_{wcs}$
    & 0.142
    & 0.395
    & 1.148
    \\ 
    $g = 0.9 g_{wcs}$
    & 0.147
    & 0.450
    & 1.448
    \\ 
    $g = 1.0 g_{wcs}$
    & 0.194
    & 0.595
    & 1.798
    \\ \hline
  \end{tabular}
  \caption{Minimum detuning between qubit frequencies, in GHz, for different combinations of acceptable average gate fidelities and coupling strengths, expressed in terms of the worst-case scenario value where the undesired gate is as strong as the desired one.}
  \label{table:single_qubit_gates}
\end{table}

\subsubsection{Colliding CZ with desired CZ in the same coupler}
\label{sec:colliding_CZ_desired_CZ_same_coupler}

\begin{itemize}
  \item Desired gate: $U_{g} = \text{CZ}$ in $t_{g} = 200 \, \mathrm{ns}$.
  \item Gate's Rabi-oscillation coupling strength: $g_{g} = 2 \pi / t_{g} = 5 \, \mathrm{MHz}$.
  \item Colliding process (assuming the desired gate is CZ02 and the colliding gate is CZ20, without loss of generality):
    \begin{align}
      H_{rf} =&\ \frac{g_{g}}{2} \lvert 02 \rangle \langle 11 \rvert + \frac{g}{2} e^{i \Delta t} \lvert 20 \rangle \langle 11 \rvert + \mathrm{H.c.}
     \end{align}
  \item Worst-case coupling strength: as strong as a normal CZ gate, $g_{wcs} = 5 \, \mathrm{MHz}$. However, we know that it will always be a fraction of this, because the collision will only be possible due to the harmonics of the low-frequency CZ. The reason for this is that the two CZ's in a given coupler lie at a distance given by the sum of the anharmonicities of the qubits involved, and the maximum acceptable detuning that we obtain even in the worst-case scenario will always be smaller than the detuning between the two gates. Note that crosstalk is irrelevant here, since these are gates on the same coupler.
  \item Detuning (collision bound) as a function of coupling strength and fidelity threshold is shown in Table~\ref{table:colliding_CZ_desired_CZ_same_coupler}.
\end{itemize}

\begin{table}[ht]
  \centering
  \begin{tabular}{|l | l l l|}
    \hline
    & $F > 0.99$ & $F > 0.999$ & $F > 0.9999$ \\\hline
    $g = 0.1 g_{wcs}$
    & $<$0.001
    & 0.007
    & 0.008
    \\ 
    $g = 0.2 g_{wcs}$
    & $<$0.001
    & 0.007
    & 0.017
    \\ 
    $g = 0.3 g_{wcs}$
    & $<$0.001
    & 0.012
    & 0.027
    \\ 
    $g = 0.4 g_{wcs}$
    & 0.007
    & 0.012
    & 0.037
    \\ 
    $g = 0.5 g_{wcs}$
    & 0.007
    & 0.017
    & 0.052
    \\ 
    $g = 0.6 g_{wcs}$
    & 0.007
    & 0.022
    & 0.067
    \\ 
    $g = 0.7 g_{wcs}$
    & 0.012
    & 0.027
    & 0.087
    \\ 
    $g = 0.8 g_{wcs}$
    & 0.012
    & 0.037
    & 0.112
    \\ 
    $g = 0.9 g_{wcs}$
    & 0.017
    & 0.042
    & 0.137
    \\ 
    $g = 1.0 g_{wcs}$
    & 0.017
    & 0.052
    & 0.167
    \\ \hline
  \end{tabular}
  \caption{Minimum detuning in GHz between the two CZ gates of the same coupler, for different combinations of acceptable average gate fidelities and coupling strengths, expressed in terms of the worst-case scenario value where the undesired CZ is as strong as the desired CZ, which is 200 ns long.}
  \label{table:colliding_CZ_desired_CZ_same_coupler}
\end{table}

\subsubsection{Colliding iSWAP with desired CZ in the same coupler}
\label{sec:colliding_iSWAP_desired_CZ_same_coupler}

\begin{itemize}
  \item Desired gate: $U_{g} = \text{CZ}$ in $t_{g} = 200 \, \mathrm{ns}$.
  \item Gate's Rabi-oscillation coupling strength: $g_{g} = 2 \pi / t_{g} = 5 \, \mathrm{MHz}$.
  \item Colliding process:
    \begin{align}
      H_{rf} =&\ \frac{g_{g}}{2} \lvert 20 / 02 \rangle \langle 11 \rvert + \frac{g}{2} e^{i \Delta t}\lvert 10 \rangle \langle 01 \rvert + \mathrm{H.c.}
    \end{align}
  \item Worst-case coupling strength: $\sqrt{2}$ weaker than the gate's coupling strength, $g_{wcs} = \sqrt{2} \pi / t_{g} \approx 3.57 \, \mathrm{MHz}$. The reason for that is that the capacitive coupling terms in the Hamiltonian ($\propto a^{\dagger} b + a b^{\dagger}$) couple states $\lvert 11 \rangle$ and $\lvert 20 \rangle$ with a strength $\sqrt{2}$ higher than states $\lvert 01 \rangle$ and $\lvert 10 \rangle$. As a result, the coupling strength of the iSWAP gate is $\sqrt{2}$ weaker, even though the gate ends up being faster due to the fact that it only entails half of the Rabi oscillation. The actual coupling strength will be approximately equal to the worst-case scenario.
  \item Detuning (collision bound) as a function of coupling strength and fidelity threshold is shown in Table~\ref{table:colliding_iSWAP_desired_CZ_same_coupler}.
\end{itemize}

\begin{table}[ht]
  \centering
  \begin{tabular}{|l | l l l|}
    \hline
    & $F > 0.99$ & $F > 0.999$ & $F > 0.9999$ \\\hline
    $g = 0.1 g_{wcs}$
    & 0.003
    & 0.005
    & 0.010
    \\ 
    $g = 0.2 g_{wcs}$
    & 0.004
    & 0.005
    & 0.010
    \\ 
    $g = 0.3 g_{wcs}$
    & 0.005
    & 0.010
    & 0.025
    \\ 
    $g = 0.4 g_{wcs}$
    & 0.005
    & 0.015
    & 0.040
    \\ 
    $g = 0.5 g_{wcs}$
    & 0.009
    & 0.020
    & 0.065
    \\ 
    $g = 0.6 g_{wcs}$
    & 0.010
    & 0.030
    & 0.090
    \\ 
    $g = 0.7 g_{wcs}$
    & 0.015
    & 0.040
    & 0.125
    \\ 
    $g = 0.8 g_{wcs}$
    & 0.019
    & 0.055
    & 0.160
    \\ 
    $g = 0.9 g_{wcs}$
    & 0.020
    & 0.065
    & 0.205
    \\ 
    $g = 1.0 g_{wcs}$
    & 0.025
    & 0.080
    & 0.250
    \\ \hline
  \end{tabular}
  \caption{Minimum detuning between iSWAP and CZ, in GHz, for different combinations of acceptable average gate fidelities and coupling strengths, expressed in terms of the worst-case scenario value where the undesired iSWAP is $\sqrt{2}$ weaker than the desired CZ, which is 200 ns long.}
  \label{table:colliding_iSWAP_desired_CZ_same_coupler}
\end{table}

\subsubsection{Colliding CZ with desired iSWAP in the same coupler}
\label{sec:colliding_CZ_desired_iSWAP_same_coupler}

\begin{itemize}
  \item Desired gate: $U_{g} = \text{iSWAP}$ in $t_{g} = 140 \, \mathrm{ns}$.
  \item Gate's Rabi-oscillation coupling strength: $g_{g} = \pi / t_{g} \approx 3.57 \, \mathrm{MHz}$.
  \item Colliding process:
    \begin{align}
      H_{rf} =&\ \frac{g_{g}}{2} \lvert 10 \rangle \langle 01 \rvert + \frac{g}{2} e^{i \Delta t} \lvert 20 / 02 \rangle \langle 11 \rvert + \mathrm{H.c.}
    \end{align}
  \item Worst-case coupling strength: $\sqrt{2}$ stronger than gate's coupling strength, $g_{wcs} = \sqrt{2} \pi / t_{g} \approx 5 \, \mathrm{MHz}$. The reasoning is the opposite as in Section \ref{sec:colliding_iSWAP_desired_CZ_same_coupler} about colliding iSWAP with desired CZ in the same coupler. The actual coupling strength will be approximately equal to the worst-case scenario.
  \item Detuning (collision bound) as a function of coupling strength and fidelity threshold is shown in Table~\ref{table:colliding_CZ_desired_iSWAP_same_coupler}.
\end{itemize}

\begin{table}[ht]
  \centering
  \begin{tabular}{|l | l l l|}
    \hline
    & $F > 0.99$ & $F > 0.999$ & $F > 0.9999$ \\\hline
    $g = 0.1 g_{wcs}$
    & $<$0.001
    & 0.006
    & 0.007
    \\ 
    $g = 0.2 g_{wcs}$
    & 0.005
    & 0.007
    & 0.014
    \\ 
    $g = 0.3 g_{wcs}$
    & 0.006
    & 0.007
    & 0.028
    \\ 
    $g = 0.4 g_{wcs}$
    & 0.006
    & 0.014
    & 0.043
    \\ 
    $g = 0.5 g_{wcs}$
    & 0.007
    & 0.021
    & 0.057
    \\ 
    $g = 0.6 g_{wcs}$
    & 0.013
    & 0.028
    & 0.086
    \\ 
    $g = 0.7 g_{wcs}$
    & 0.013
    & 0.036
    & 0.107
    \\ 
    $g = 0.8 g_{wcs}$
    & 0.014
    & 0.050
    & 0.143
    \\ 
    $g = 0.9 g_{wcs}$
    & 0.020
    & 0.057
    & 0.186
    \\ 
    $g = 1.0 g_{wcs}$
    & 0.027
    & 0.071
    & 0.221
    \\ \hline
  \end{tabular}
  \caption{Minimum detuning between iSWAP and CZ, in GHz, for different combinations of acceptable average gate fidelities and coupling strengths, expressed in terms of the worst-case scenario value where the undesired CZ is $\sqrt{2}$ stronger than the desired iSWAP, which is 140 ns long.}
  \label{table:colliding_CZ_desired_iSWAP_same_coupler}
\end{table}

\subsubsection{Colliding CZ with desired CZ in neighbor coupler without sharing other state than $\lvert 111 \rangle$}
\label{sec:colliding_CZ_desired_CZ_neighbor_coupler}

\begin{itemize}
  \item Desired gate: $U_{g} = \text{CZ} \otimes \mathbb{I}$ in $t_{g} = 200 \, \mathrm{ns}$.
  \item Gate's Rabi-oscillation coupling strength: $g_{g} = 2 \pi / t_{g} = 5 \, \mathrm{MHz}$.
  \item Colliding process (assuming the desired gate is CZ20 and the colliding gate is CZ20):
    \begin{align}
      H_{rf} =&\ \frac{g_{g}}{2} \left( \lvert 200 \rangle \langle 110 \rvert + \lvert 201 \rangle \langle 111 \rvert \right) \nonumber \\
      +&\ \frac{g}{2} e^{i \Delta t} \left( \lvert 020 \rangle \langle \lvert 011 \rvert + \lvert 120 \rangle \langle 111 \rvert \right) + \mathrm{H.c.}
    \end{align}
    This section also covers the two following collisions:
    \begin{itemize}
      \item CZ20 $\leftarrow$ CZ02. It requires the substitution of states $\lvert 020 \rangle$ and $\lvert 120 \rangle$ by $\lvert 002 \rangle$ and $\lvert 102 \rangle$, respectively.
      \item CZ02 $\leftarrow$ CZ02. It requires the substitution of states $\lvert 020 \rangle$, $\lvert 120 \rangle$, $\lvert 200 \rangle$ and $\lvert 201 \rangle$ by $\lvert 002 \rangle$, $\lvert 102 \rangle$, $\lvert 020 \rangle$ and $\lvert 021$, respectively.
    \end{itemize}
  \item Worst-case coupling strength: as strong as the gate's coupling strength, $g_{wcs} = 5 \, \mathrm{MHz}$. As the process will be applied due to crosstalk, the actual coupling will be a fraction of this.
  \item Detuning (collision bound) as a function of coupling strength and fidelity threshold is shown in Table~\ref{table:colliding_CZ_desired_CZ_neighbor_coupler}.
\end{itemize}

\begin{table}[ht]
  \centering
  \begin{tabular}{|l | l l l|}
    \hline
    & $F > 0.99$ & $F > 0.999$ & $F > 0.9999$ \\\hline
    $g = 0.1 g_{wcs}$
    & 0.002
    & 0.007
    & 0.017
    \\ 
    $g = 0.2 g_{wcs}$
    & 0.005
    & 0.012
    & 0.036
    \\ 
    $g = 0.3 g_{wcs}$
    & 0.006
    & 0.018
    & 0.056
    \\ 
    $g = 0.4 g_{wcs}$
    & 0.008
    & 0.025
    & 0.079
    \\ 
    $g = 0.5 g_{wcs}$
    & 0.011
    & 0.034
    & 0.106
    \\ 
    $g = 0.6 g_{wcs}$
    & 0.014
    & 0.044
    & 0.137
    \\ 
    $g = 0.7 g_{wcs}$
    & 0.017
    & 0.055
    & 0.172
    \\ 
    $g = 0.8 g_{wcs}$
    & 0.021
    & 0.067
    & 0.213
    \\ 
    $g = 0.9 g_{wcs}$
    & 0.026
    & 0.082
    & 0.257
    \\ 
    $g = 1.0 g_{wcs}$
    & 0.031
    & 0.097
    & 0.307
    \\ \hline
  \end{tabular}
  \caption{Minimum detuning between CZs associated to neighboring couplers, in GHz, for different combinations of acceptable average gate fidelities and coupling strengths, expressed in terms of the worst-case scenario value where the undesired CZ is as strong as the desired CZ, which is 200 ns long.}
  \label{table:colliding_CZ_desired_CZ_neighbor_coupler}
\end{table}

\subsubsection{Colliding CZ with desired CZ in neighbor coupler sharing other states than $\lvert 111 \rangle$}
\label{sec:colliding_CZ_desired_CZ_neighbor_coupler_more_states}

\begin{itemize}
  \item Desired gate: $U_{g} = \text{CZ} \otimes \mathbb{I}$ in $t_{g} = 200 \, \mathrm{ns}$.
  \item Gate's Rabi-oscillation coupling strength: $g_{g} = 2 \pi / t_{g} = 5 \, \mathrm{MHz}$.
  \item Colliding process:
    \begin{align}
      H_{rf} =&\ \frac{g_{g}}{2} \left( \lvert 020 \rangle \langle 110 \rvert + \lvert 201 \rangle \langle \langle 111 \rvert\right) \nonumber \\
      +& \frac{g}{2} e^{i \Delta t} \left( \lvert 020 \rangle \langle 011 \rvert + \lvert 120 \rangle \langle 111 \rvert \right) + \mathrm{H.c.}
    \end{align}
  \item Worst-case coupling strength: as strong as the gate's coupling strength, $g_{wcs} = 5 \, \mathrm{MHz}$. As the process will be applied due to crosstalk, the actual coupling will be a small fraction of this.
  \item Detuning (collision bound) as a function of coupling strength and fidelity threshold is shown Table~\ref{table:colliding_CZ_desired_CZ_neighbor_coupler_more_states}.
\end{itemize}

\begin{table}[ht]
  \centering
  \begin{tabular}{|l | l l l|}
    \hline
    & $F > 0.99$ & $F > 0.999$ & $F > 0.9999$ \\\hline
    $g = 0.1 g_{wcs}$
    & $<$0.001
    & 0.007
    & 0.012
    \\ 
    $g = 0.2 g_{wcs}$
    & 0.006
    & 0.007
    & 0.022
    \\ 
    $g = 0.3 g_{wcs}$
    & 0.007
    & 0.012
    & 0.037
    \\ 
    $g = 0.4 g_{wcs}$
    & 0.007
    & 0.022
    & 0.057
    \\ 
    $g = 0.5 g_{wcs}$
    & 0.012
    & 0.027
    & 0.087
    \\ 
    $g = 0.6 g_{wcs}$
    & 0.012
    & 0.042
    & 0.122
    \\ 
    $g = 0.7 g_{wcs}$
    & 0.017
    & 0.052
    & 0.162
    \\ 
    $g = 0.8 g_{wcs}$
    & 0.022
    & 0.067
    & 0.207
    \\ 
    $g = 0.9 g_{wcs}$
    & 0.027
    & 0.082
    & 0.257
    \\ 
    $g = 1.0 g_{wcs}$
    & 0.032
    & 0.102
    & 0.317
    \\ \hline
  \end{tabular}
  \caption{Minimum detuning between CZs associated to neighboring couplers, in GHz, for different combinations of acceptable average gate fidelities and coupling strengths, expressed in terms of the worst-case scenario value where the undesired CZ is as strong as the desired CZ, which is 200 ns long.}
  \label{table:colliding_CZ_desired_CZ_neighbor_coupler_more_states}
\end{table}

\subsubsection{Colliding iSWAP with desired CZ in neighbor coupler}
\label{sec:colliding_iSWAP_desired_CZ_neighbor_coupler}

\begin{itemize}
  \item Desired gate: $U_{g} = \text{CZ} \otimes \mathbb{I}$, in $t_{g} = 200 \, \mathrm{ns}$.
  \item Gate's Rabi-oscillation coupling strength: $g_{g} = 2 \pi / t_{g} = 5 \, \mathrm{MHz}$.
  \item Colliding process (assuming CZ20, without loss of generality):
    \begin{align}
      H_{rf} =&\ \frac{g_{g}}{2} \left( \lvert 200 \rangle \langle 110 \rvert + \lvert 201 \rangle \langle 111 \rvert \right) \nonumber \\
      +&\ \frac{g}{2} e^{i \Delta t} \left( \lvert 010 \rangle \langle 001 \rvert + \lvert 110 \rangle \langle 101 \rvert + \lvert 210 \rangle \langle 201 \rvert \right) + \mathrm{H.c.}
    \end{align}
    If the desired gate were CZ02, the states involved would change from $\lvert 200 \rangle$, $\lvert 201 \rangle$ and $\lvert 210 \rangle$ to $\lvert 020 \rangle$, $\lvert 021 \rangle$ and $\lvert 201 \rangle$, respectively.
  \item Worst-case coupling strength: $\sqrt{2}$ weaker than the gate's coupling strength, $g_{wcs} = \sqrt{2} \pi / t_{g} \approx 3.5 \, \mathrm{MHz}$, because that is how strong iSWAPs can roughly be. It will actually be a fraction of that because the process is activated due to crosstalk.
  \item Detuning (collision bound) as a function of coupling strength and fidelity threshold is shown Table~\ref{table:colliding_iSWAP_desired_CZ_neighbor_coupler}.
\end{itemize}

\begin{table}[ht]
  \centering
  \begin{tabular}{|l | l l l|}
    \hline
    & $F > 0.99$ & $F > 0.999$ & $F > 0.9999$ \\\hline
    $g = 0.1 g_{wcs}$
    & 0.003
    & 0.007
    & 0.017
    \\ 
    $g = 0.2 g_{wcs}$
    & 0.006
    & 0.012
    & 0.035
    \\ 
    $g = 0.3 g_{wcs}$
    & 0.007
    & 0.018
    & 0.055
    \\ 
    $g = 0.4 g_{wcs}$
    & 0.008
    & 0.025
    & 0.077
    \\ 
    $g = 0.5 g_{wcs}$
    & 0.011
    & 0.033
    & 0.103
    \\ 
    $g = 0.6 g_{wcs}$
    & 0.014
    & 0.042
    & 0.133
    \\ 
    $g = 0.7 g_{wcs}$
    & 0.017
    & 0.053
    & 0.166
    \\ 
    $g = 0.8 g_{wcs}$
    & 0.021
    & 0.065
    & 0.204
    \\ 
    $g = 0.9 g_{wcs}$
    & 0.025
    & 0.078
    & 0.246
    \\ 
    $g = 1.0 g_{wcs}$
    & 0.03
    & 0.093
    & 0.293
    \\ \hline
  \end{tabular}
  \caption{Minimum detuning between CZ and iSWAP in neighboring couplers, in GHz, for different combinations of acceptable average gate fidelities and coupling strengths, expressed in terms of the worst-case scenario value where the undesired iSWAP is $\sqrt{2}$ weaker than the desired CZ, which is 200 ns long.}
  \label{table:colliding_iSWAP_desired_CZ_neighbor_coupler}
\end{table}

\subsubsection{Colliding CZ with desired iSWAP in neighbor coupler without sharing other state than $\lvert 011 \rangle$}
\label{sec:colliding_CZ_desired_iSWAP_neighbor_coupler}

\begin{itemize}
  \item Desired gate: $U_{g} = \text{iSWAP} \otimes \mathbb{I}$, in $t_{g} = 140 \, \mathrm{ns}$.
  \item Gate's Rabi-oscillation coupling strength: $g_{g} = \pi / t_{g} \approx 3.5 \, \mathrm{MHz}$.
  \item Colliding process:
    \begin{align}
      H_{rf} =&\ \frac{g_{g}}{2} \left( \lvert 100 \rangle \langle 010 \rvert + \lvert 101 \rangle \langle 011 \rvert \right) \nonumber \\
      +&\ \frac{g}{2} e^{i \Delta t} \left( \lvert 020 \rangle \langle 011 \rvert + \lvert 120 \rangle \langle 111 \rvert \right) + \mathrm{H.c.}
    \end{align}
  \item Worst-case coupling strength: $\sqrt{2}$ stronger than the gate's coupling strength, $g_{wcs} = \sqrt{2} \pi / t_{g} \approx 5 \, \mathrm{MHz}$, because that is how strong CZs can roughly be. It will actually be a fraction of that because the process is activated due to crosstalk.
  \item Detuning (collision bound) as a function of coupling strength and fidelity threshold is shown in Table~\ref{table:colliding_CZ_desired_iSWAP_neighbor_coupler}.
\end{itemize}

\begin{table}[ht]
  \centering
  \begin{tabular}{|l | l l l|}
    \hline
    & $F > 0.99$ & $F > 0.999$ & $F > 0.9999$ \\\hline
    $g = 0.1 g_{wcs}$
    & 0.0
    & 0.006
    & 0.008
    \\ 
    $g = 0.2 g_{wcs}$
    & 0.005
    & 0.007
    & 0.021
    \\ 
    $g = 0.3 g_{wcs}$
    & 0.006
    & 0.013
    & 0.029
    \\ 
    $g = 0.4 g_{wcs}$
    & 0.006
    & 0.014
    & 0.043
    \\ 
    $g = 0.5 g_{wcs}$
    & 0.007
    & 0.021
    & 0.064
    \\ 
    $g = 0.6 g_{wcs}$
    & 0.012
    & 0.028
    & 0.086
    \\ 
    $g = 0.7 g_{wcs}$
    & 0.013
    & 0.036
    & 0.114
    \\ 
    $g = 0.8 g_{wcs}$
    & 0.014
    & 0.049
    & 0.143
    \\ 
    $g = 0.9 g_{wcs}$
    & 0.02
    & 0.057
    & 0.178
    \\ 
    $g = 1.0 g_{wcs}$
    & 0.021
    & 0.071
    & 0.214
    \\ \hline
  \end{tabular}
  \caption{Minimum detuning between CZ and iSWAP in neighboring couplers, in GHz, for different combinations of acceptable average gate fidelities and coupling strengths, expressed in terms of the worst-case scenario value where the undesired CZ is $\sqrt{2}$ stronger than the desired iSWAP, which is 140 ns long.}
  \label{table:colliding_CZ_desired_iSWAP_neighbor_coupler}
\end{table}

\subsubsection{Colliding CZ with desired iSWAP in neighbor coupler sharing other states than $\lvert 011 \rangle$}
\label{sec:colliding_CZ_desired_iSWAP_neighbor_coupler_more_states}

\begin{itemize}
  \item Desired gate: $U_{g} = \text{iSWAP} \otimes \mathbb{I}$, in $t_{g} = 140 \, \mathrm{ns}$.
  \item Gate's Rabi-oscillation coupling strength: $g_{g} = \pi / t_{g} \approx 3.5 \, \mathrm{MHz}$.
  \item Colliding process:
    \begin{align}
      H_{rf} =&\ \frac{g_{g}}{2} \left( \lvert 100 \rangle \langle 010 \rvert + \lvert 101 \rangle \langle 011 \rvert + \lvert 012 \rangle \langle 102 \rvert \right) \nonumber \\ 
      +& \frac{g}{2} e^{i \Delta t} \left( \lvert 002 \rangle \langle 011 \rvert + \lvert 102 \rangle \langle 111 \rvert \right) + \mathrm{H.c.}
    \end{align}
  \item Worst-case coupling strength: $\sqrt{2}$ stronger than the gate's coupling strength, $g_{wcs} = \sqrt{2} \pi / t_{g} \approx 5 \, \mathrm{MHz}$, because that's how strong CZ's can roughly be. It will actually be a fraction of that because the process is activated due to crosstalk.
  \item Detuning (collision bound) as a function of coupling strength and fidelity threshold is shown in Table~\ref{table:colliding_CZ_desired_iSWAP_neighbor_coupler_more_states}.
\end{itemize}

\begin{table}[ht]
  \centering
  \begin{tabular}{|l | l l l|}
    \hline
    & $F > 0.99$ & $F > 0.999$ & $F > 0.9999$ \\\hline
    $g = 0.1 g_{wcs}$
    & $<$0.001
    & 0.007
    & 0.015
    \\ 
    $g = 0.2 g_{wcs}$
    & 0.005
    & 0.008
    & 0.029
    \\ 
    $g = 0.3 g_{wcs}$
    & 0.007
    & 0.015
    & 0.044
    \\ 
    $g = 0.4 g_{wcs}$
    & 0.007
    & 0.021
    & 0.064
    \\ 
    $g = 0.5 g_{wcs}$
    & 0.008
    & 0.028
    & 0.085
    \\ 
    $g = 0.6 g_{wcs}$
    & 0.013
    & 0.035
    & 0.108
    \\ 
    $g = 0.7 g_{wcs}$
    & 0.014
    & 0.043
    & 0.136
    \\ 
    $g = 0.8 g_{wcs}$
    & 0.016
    & 0.052
    & 0.167
    \\ 
    $g = 0.9 g_{wcs}$
    & 0.021
    & 0.065
    & 0.203
    \\ 
    $g = 1.0 g_{wcs}$
    & 0.024
    & 0.078
    & 0.243
    \\ \hline
  \end{tabular}
  \caption{Minimum detuning between CZ and iSWAP in neighboring couplers, in GHz, for different combinations of acceptable average gate fidelities and coupling strengths, expressed in terms of the worst-case scenario value where the undesired CZ is $\sqrt{2}$ stronger than the desired iSWAP, which is 140 ns long.}
  \label{table:colliding_CZ_desired_iSWAP_neighbor_coupler_more_states}
\end{table}

\subsubsection{Colliding iSWAP with desired iSWAP in neighbor coupler}
\label{sec:colliding_iSWAP_desired_iSWAP_neighbor_coupler}

\begin{itemize}
  \item Desired gate: $U_{g} = \text{iSWAP} \otimes \mathbb{I}$, in $t_{g} = 140 \, \mathrm{ns}$.
  \item Gate's Rabi-oscillation coupling strength: $g_{g} = \pi / t_{g} \approx 3.5 \, \mathrm{MHz}$.
  \item Colliding process:
    \begin{align}
      H_{rf} =&\ \frac{g_{g}}{2} \left( \lvert 100 \rangle \langle 010 \rvert + \lvert 101 \rangle \langle 011 \rvert \right) \nonumber \\
      +&\ \frac{g}{2} e^{i \Delta t} \left( \lvert 010 \rangle \langle 001 \rvert + \lvert 110 \rangle \langle 101 \rvert \right) + \mathrm{H.c.}
    \end{align}
  \item Worst-case coupling strength: as strong as the gate's coupling strength, $g_{wcs} = \pi / t_{g} \approx 3.5 \, \mathrm{MHz}$, because that is how strong iSWAPs can roughly be. It will actually be a fraction of that because the process is activated due to crosstalk.
  \item Detuning (collision bound) as a function of coupling strength and fidelity threshold is shown in Table~\ref{table:colliding_iSWAP_desired_iSWAP_neighbor_coupler}.
\end{itemize}

\begin{table}[ht]
  \centering
  \begin{tabular}{|l | l l l|}
    \hline
    & $F > 0.99$ & $F > 0.999$ & $F > 0.9999$ \\\hline
    $g = 0.1 g_{wcs}$
    & $<$0.001
    & 0.007
    & 0.016
    \\ 
    $g = 0.2 g_{wcs}$
    & 0.006
    & 0.013
    & 0.036
    \\ 
    $g = 0.3 g_{wcs}$
    & 0.007
    & 0.016
    & 0.052
    \\ 
    $g = 0.4 g_{wcs}$
    & 0.008
    & 0.023
    & 0.073
    \\ 
    $g = 0.5 g_{wcs}$
    & 0.009
    & 0.030
    & 0.094
    \\ 
    $g = 0.6 g_{wcs}$
    & 0.013
    & 0.037
    & 0.120
    \\ 
    $g = 0.7 g_{wcs}$
    & 0.015
    & 0.045
    & 0.145
    \\ 
    $g = 0.8 g_{wcs}$
    & 0.017
    & 0.056
    & 0.174
    \\ 
    $g = 0.9 g_{wcs}$
    & 0.021
    & 0.065
    & 0.207
    \\ 
    $g = 1.0 g_{wcs}$
    & 0.024
    & 0.077
    & 0.242
    \\ \hline
  \end{tabular}
  \caption{Minimum detuning between iSWAP's in neighboring couplers, in GHz, for different combinations of acceptable average gate fidelities and coupling strengths, expressed in terms of the worst-case scenario value where the undesired iSWAP is as strong as the desired iSWAP, which is 140 ns long.}
  \label{table:colliding_iSWAP_desired_iSWAP_neighbor_coupler}
\end{table}

\vspace{2cm}
\newpage
\bibliography{reference_PRXQ}

\end{document}